\documentclass[noeprint,twocolumn,showpacs,amsmath,amssymb,sort&compress,floatfix,superscriptaddress,pre,aps]{revtex4-2}

\usepackage{graphicx,color}
\usepackage{dcolumn}
\usepackage{xcolor}
\usepackage[normalem]{ulem}
\usepackage{bm}
\usepackage[hypertexnames=false]{hyperref}
\usepackage[mathlines]{lineno}
\usepackage{mathtools}
\usepackage{amsmath}
\usepackage{float}
\usepackage[caption=false]{subfig}
\usepackage[english]{babel}
\usepackage{xcolor}
\usepackage{xparse}
\addto\captionsenglish{}
\usepackage{flushend}
\usepackage{nowidow}
\usepackage{balance}
\usepackage{placeins}

\newcommand{\ucsb}{Department of Physics, University of California Santa Barbara, Santa Barbara, CA 93106, USA}
\newcommand{\ue}{SUPA, School of Physics and Astronomy, University of Edinburgh, Peter Guthrie Tait Road, Edinburgh EH9 3FD, United Kingdom}
\newcommand{\uc}{Facultad de F{\'i}sica, Pontificia Universidad Cat{\'o}lica de Chile, Santiago 7820436, Chile}

\newcommand{\rx}{\mathrm x}
\newcommand{\rv}{\mathrm v}
\newcommand{\rp}{\mathrm p}
\newcommand{\rt}{\mathrm t}
\newcommand{\rf}{\mathrm f}
\newcommand{\rF}{\mathrm F}
\newcommand{\rX}{\mathrm X}
\newcommand{\rV}{\mathrm V}
\newcommand{\rP}{\mathrm P}
\newcommand{\rT}{\mathrm T}

\newcommand{\bx}{\bm\rx}
\newcommand{\bv}{\bm\rv}
\newcommand{\bp}{\bm\rp}
\newcommand{\bt}{\bm\rt}
\newcommand{\bff}{\bm\rf}
\newcommand{\bF}{\bm\rF}
\newcommand{\bX}{\bm\rX}
\newcommand{\bV}{\bm\rV}
\newcommand{\bP}{\bm\rP}
\newcommand{\bT}{\bm\rT}

\newcommand{\F}{\mathcal{F}}

\renewcommand{\vec}[1]{\bm{#1}}

\newcommand{\trace}{\textrm{Tr}}
\renewcommand{\O}{\mathcal{O}}
\newcommand{\fd}{\textrm{d}}
\newcommand{\pd}{\partial}

\newcommand{\oavg}[1]{\overline{#1}}
\newcommand{\vepsi}{\varepsilon}

\newcommand{\icmp}{\perp}
\newcommand{\ocmp}{z}

\newcommand{\eqn}{Eq.~}
\newcommand{\eqns}{Eqs.~}
\newcommand{\fig}{Fig.~}
\newcommand{\figs}{Figs.~}

\newcommand{\app}{Appendix~}

\begin{document}

\title{Multiphase Field Model of Cells on a Substrate: From 3D to 2D}

\author{Michael Chiang}
\affiliation{\ue}
\author{Austin Hopkins}
\affiliation{\ucsb}
\author{Benjamin Loewe}
\affiliation{\ue}
\affiliation{\uc}
\author{Davide Marenduzzo}
\affiliation{\ue}
\author{M. Cristina Marchetti}
\affiliation{\ucsb}

\date{\today}

\begin{abstract}
Multiphase field models have emerged as an important computational tool for understanding biological tissue while resolving single-cell properties. While they have successfully reproduced many experimentally observed behaviors of living tissue, the theoretical underpinnings have not been fully explored. We show that a two-dimensional version of the model, which is commonly employed to study tissue monolayers, can be derived from a three-dimensional version in the presence of a substrate. We also show how viscous forces, which arise from friction between different cells, can be included in the model. Finally, we numerically simulate a tissue monolayer, and find that intercellular friction tends to solidify the tissue.
\end{abstract}

\maketitle

\section{Introduction}

Phase field models have been used extensively to describe the behavior of monolayers of living cells \cite{shao2010computational,nonomura2012study,ziebert2012model,palmieri2015multiple,najem2016phase,marth2016collective,moure2016computational,mueller2019emergence,yang2019computational,alert2020physical,wenzel2021multiphase}. In these models, each cell is described by a scalar field defined to be unity where the cell is and zero outside. The dynamics of the field is governed by thermodynamic forces controlled by gradients in the local chemical potential and advection by the cell's own velocity, which in turn arises from cell motility and interactions with other cells. The phase field model is highly versatile, as it allows the inclusion of repulsive and adhesive cell-cell interactions, cell contractility and motility, dissipative forces due to intercellular friction and friction with a substrate, and cell division and apoptosis. In contrast to vertex (e.g.,~\cite{honda1983geometrical,nagai2001dynamic,staple2010mechanics,fletcher2014vertex,bi2015density}) and Voronoi (e.g.,~\cite{li2014coherent,bi2016motility}) models, which are generally restricted to confluent tissue, with no gaps between cells (although versions applicable to non-confluent situations have been implemented~\cite{kim2021embryonic, huang2023bridging}), it naturally allows both variations in cell shape and cell density. Phase field models have been shown to capture several experimentally observed behaviors, including solid-liquid transitions in dense biological tissue~\cite{mueller2019emergence,balasubramaniam2021investigating}, the emergence of nematic~\cite{mueller2019emergence} and hexatic order in epithelial layers~\cite{armengol2023epithelia}, and the correlation between topological defects in the tissue structure and biological functions, such as cell extrusion and apoptosis~\cite{monfared2023mechanical}. Extensions to three dimensions are also being implemented~\cite{monfared2023mechanical,kuang2023morphosim}.

An important component of the phase field model is the free energy functional that governs the shape of the phase field representing a cell. Processes such as self-propulsion and collisions with other cells can drive deformations of a cell from its preferred shape, which incur a free energy penalty, resulting into a passive thermodynamic (or capillary) stress~\cite{cates2018theories}. Different realizations of the model have either included~\cite{palmieri2015multiple,mueller2019emergence,peyret2019sustained,zhang2020active,balasubramaniam2021investigating,zadeh2022picking, zhang2023active,monfared2023mechanical,hopkins2023motility,chiang2023intercellular} or neglected \cite{loewe2020solid,hopkins2022local,armengol2023epithelia} the resulting forces in the force balance equation, and it remains unclear whether the inclusion of such passive stress in controlling the cell advection velocity is crucial in determining the collective dynamics of the tissue. Finally, most studies start with a set of particle-like, two-dimensional (2D) equations for individual phase fields focusing on the in-plane dynamics of the cells on the substrate, and a theory that explicitly links the full 3D phase field dynamics of the tissue to this effective 2D dynamics is still lacking.

In this paper, we derive a set of effective 2D equations of motion for the cells by starting from a model of interacting cells in 3D and then averaging out the dynamics in the direction normal to the substrate, in the limit where the height of each cell is small compared to its lateral extent~\cite{banerjee2011substrate,banerjee2012contractile}. This derivation shows that passive capillary stress naturally arises in the force balance equation from the interactions of cells with a frictional substrate. Using simulations, we further demonstrate that these passive stresses play a significant role in controlling the rheological state of the tissue. A new element of our work is the inclusion of cell-cell friction which leads to viscous-type forces at the tissue scale and proves to also have an important effect on the solid-liquid transition. The method considered here is analogous to the lubrication approximation commonly used in fluid dynamics to derive equations for a thin liquid film~\cite{oron1997long}.

The rest of the paper is organized as follows. In Section II, we develop the theoretical framework for averaging the out-of-plane dynamics by considering a simpler system with a single cell located on the substrate. We show how one can derive a 2D particle-like description of the cell, with its kinematic properties, such as its velocity and self-propulsion polarity, arising from coarse-grained body forces (and captured by coarse-grained body velocity vectors). In Section III, we extend this framework to modeling multiple cells on the substrate. In particular, we incorporate viscous stress within the 3D tissue and demonstrate that it encapsulates pairwise friction between cells in the projected 2D dynamics. Section IV provides simulation results from solving the projected 2D equations of motion for the cells. Notably, we find that passive stresses strongly influence collective motion, for instance, in altering the transition boundary of the solid-liquid transition driven by cell deformability and activity. Section V compares the effect of energetic adhesive interactions between cells and intercellular friction on tissue dynamics, and we observe that the latter contributes more significantly to the correlations in cell velocities. Finally, in Section VI we summarize the results and discuss implications of this work.

\section{Model for a single cell}

We begin by examining the case of a single cell situated on a planar rigid substrate. This allows one to familiarize with the phase field approach in modeling cells and develop the fundamental framework for deriving the effective 2D dynamics of the cell, which can then be generalized to multiple cells. Our goal is to apply the thin-layer approximation to reduce the cell dynamics from 3D to 2D and to derive a particle-like equation of motion for the phase field, whereby the cell, treated as a deformable particle, is advected by a coarse-grained body velocity vector instead of a field. As shown later, this provides a computationally tractable framework for incorporating intercellular friction when there are multiple cells.

\subsection{3D free energy and force balance}

\begin{figure}[t]
  \includegraphics[width=0.45\textwidth]{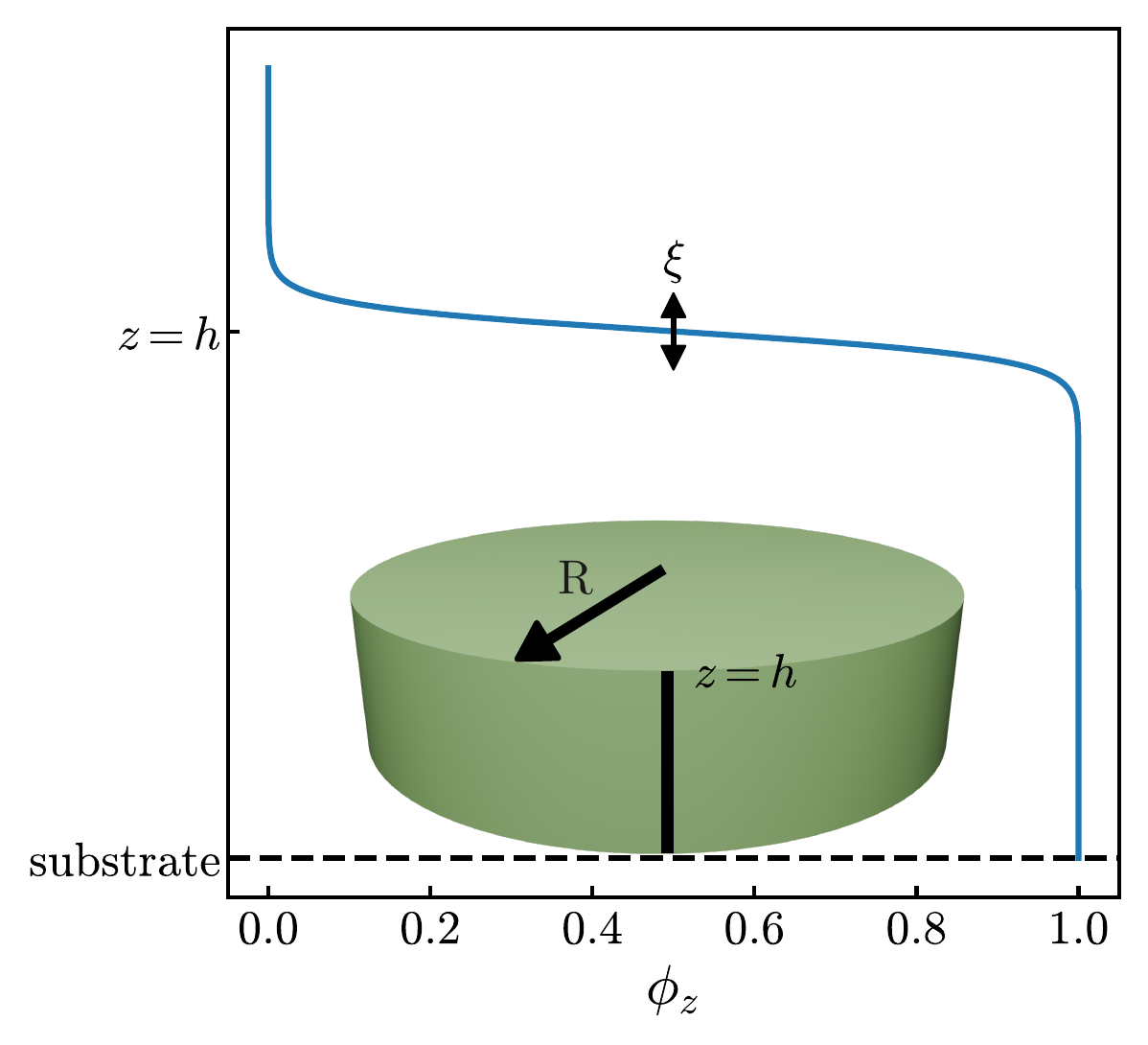}
  \caption{Illustration of the assumed profile for $\phi_{\ocmp}$, as in \eqn\eqref{eqn:phi_ocmp}, where $\phi_{\ocmp}$ approaches 1 on the substrate ($z=0$) and transitions from 1 to 0 at a height $h$ over a small width $\xi$. Inset: Illustration of the model cell in 3D.}
  \label{fig:setup}
\end{figure}

We model the cell to be in the positive half-plane, described in Cartesian coordinates $\bX=(\bx,z)$, where $z \in \left[0,h\right]$ and the substrate is at $z = 0$ (see \fig\ref{fig:setup}). The cell has a height $h$ and an interfacial thickness $\xi$. Periodic boundary conditions are assumed in the $x$-$y$ plane for convenience, without loss of generality. We describe the cell using a phase field $\phi(\bX)$, where $\phi \approx 1$ for regions within the cell and $\phi \approx 0$ for those outside the cell. This phase field represents the local concentration of cellular content (i.e., active matter and water).

For the phase field to capture the morphology of a cell, we impose a free energy $\F$ with two contributions. First, there is a Cahn-Hilliard-type free energy given by
\begin{align}
  \F^{\text{CH}} = \int\fd\bX\, f^{\text{CH}} = \kappa\int\fd\bX\,\left[f(\phi)+\xi^2(\vec{\nabla}\phi)^2\right]\;, 
  \label{eqn:3D_free_energy_CH_single}
\end{align}
in which
\begin{equation}
  f(\phi) = \phi^2(\phi-1)^2
\end{equation}
favors $\phi$ close to $0$ or $1$ when minimizing $\F^{\text{CH}}$. The second term in~\eqn\eqref{eqn:3D_free_energy_CH_single} penalizes the formation of interfaces and ensures the maintenance of a single coherent droplet, with $\xi$ a length scale that controls the thickness of the interfacial region. The edge tension $\gamma$ of the cell (related to the excess free energy to form the interface) can be written as $\gamma = \kappa\xi/3$.

Second, we assume that the total cell volume $V$ is not strictly conserved but fluctuates around a preferred value $V_0$ (e.g., water can leave or enter the cell). This is incorporated as a soft constraint in the free energy as
\begin{align}
  \F^{V} = \lambda V_0\left(1-\frac{V}{V_0}\right)^2 \equiv \lambda V_0\,\delta V^2\;,
\end{align}
where we defined the cell volume to be
\begin{equation}
  V[\phi] = \int\fd\bX\,\phi^2 \;.
\end{equation}
The choice of defining the volume with $\phi^2$ instead of $\phi$ does not qualitatively change the computed volume, as long as the interface width is small relative to the cell radius. However, it results in a volume constraint contribution to the chemical potential which depends on $\phi$, and hence gives a more physical evolution equation for the cell's phase field.

We account for the effect of mechanical forces acting on the cell using the Cauchy momentum equation. In the limit where the system is overdamped and body forces are absent, this equation becomes
\begin{equation}   
  \nabla_{\beta}\Sigma_{\alpha\beta} = 0 \;,
  \label{eqn:3D_force_balance_single}
\end{equation}
which is a statement of force balance and momentum conservation, where $\Sigma_{\alpha\beta}$ are components of the total 3D stress tensor of the cell (we use Greek indices for Cartesian components in 3D and Latin indices for those in 2D and observe Einstein summation convention). In this work, since we are interested in the 2D in-plane dynamics of the cell, we only focus on these components of this set of equations (i.e., $\alpha = a \in \{x,y\}$). The boundary conditions we assume here are that $\Sigma_{az}(\bx,z=0) \neq 0$, as the basal layer of the cell interacts with the substrate (i.e., there is momentum exchange), and $\Sigma_{az}(\bx,z=h) \approx 0$, which imposes a free boundary at the cell's apical surface.

We consider two types of stress acting on each cell element:
\begin{equation}
  \Sigma_{\alpha\beta} = \Sigma_{\alpha\beta}^{\text{pas}} + \Sigma_{\alpha\beta}^{\text{pol}} \;.
\end{equation}
Here, $\Sigma_{\alpha\beta}^{\text{pas}}$ is a passive thermodynamic (or capillary) stress that arises from cell deformation and is governed by the relation~\cite{cates2018theories}
\begin{equation}
  \nabla_{\beta}\Sigma_{\alpha\beta}^{\text{pas}} = -\phi\nabla_{\alpha}\mu \;,
  \label{eqn:3D_div_passive_stress_single}
\end{equation}
where $\mu = \delta\F/\delta\phi$ is the exchange chemical potential, given by
\begin{equation}
  \mu = 2\kappa\left[\phi(\phi-1)(2\phi-1)-\xi^2\nabla^2\phi\right]-4 \lambda \delta V \phi\;.
\end{equation}
With some algebra, one can further verify that
\begin{equation}
  \Sigma_{\alpha\beta}^{\text{pas}} = -\delta_{\alpha\beta}\Pi - 2\kappa\xi^2(\nabla_{\alpha}\phi)(\nabla_{\beta}\phi) \;,
  \label{eqn:3D_passive_stress_single}
\end{equation}
where $\Pi = \mu\phi-f^{\text{CH}}+2\lambda\delta V\phi^2$ is an osmotic pressure (see \app\ref{app:passive_stress}).

The other stress $\Sigma_{\alpha\beta}^{\text{pol}}$ accounts for active processes that drive cell motility. The self-propulsion of a cell is largely mediated by traction forces transmitted to the cell-substrate interface involving actin-based protrusions~\cite{alert2020physical}. For simplicity, we choose to model this phenomenologically, with details presented in the next section. We note that one can also incorporate other processes such as active cell contractility into the force-balance equation, which will not change the 3D-to-2D projection framework presented here.

\subsection{From 3D to 2D dynamics}

To derive a 2D particle-like model for the cell moving on a substrate from the full 3D description, we make the following approximations. First, we assume a separation of length scales: the interface width $\xi$ is much smaller than the height $h$ of the cell in the $z$ direction, which in turn is smaller than the cell's typical lateral extension $\ell$ -- i.e., $\xi \ll$ $h \ll \ell$. Second, we neglect spatial variations of the frictional forces with the substrate and of the active forces that drive cell motility over the area of the cell and define single degrees of freedom describing motility and propulsion for a cell. This will become important in developing a computationally tractable model of many interacting cells. We further assume that the phase field can be separated into the product of in-plane ($\icmp$) and out-of-plane ($\ocmp$) fields as
\begin{equation}
  \phi(\bX) = \phi_{\icmp}(\bx)\,\phi_{\ocmp}(z) \;. 
\end{equation}
This factorization facilitates the operation of averaging the phase field over the $z$ direction. Under this assumption, the cell's cross-sectional shape is independent of $z$, though it can evolve over time. Inspired by the phase field profile for a binary fluid, we assume the cell's vertical profile takes the form
\begin{equation}
  \phi_{\ocmp}(z) = \frac{1}{2}\left[1+\tanh\left(\frac{h-z}{\xi}\right)\right],
  \label{eqn:phi_ocmp}
\end{equation}
which is a smoothed step-wise function (see \fig\ref{fig:setup}).

We now proceed to obtain an in-plane-only description of the model by averaging various quantities $q$ over the $z$ direction. More precisely, we define this $z$-average operation as
\begin{equation}
  \oavg{q}(\bx) \equiv \frac{1}{h}\int_0^{h}\fd z\, q(\bX) \;.
\end{equation}
Taking the $z$-average of the force balance equation [\eqn\eqref{eqn:3D_force_balance_single}] and applying the boundary conditions lead to
\begin{equation}  
  \nabla_{b}{\sigma_{ab}} = \mathrm{t}_a(\mathbf{x})\;,
  \label{eqn:2D_traction_force_balance_single}
\end{equation}
where we defined the 2D stress tensor
\begin{equation}
  \sigma_{ab} \equiv h\oavg{\Sigma}_{ab} \;,
\end{equation}
and 
\begin{equation}
  \mathrm{t}_a(\bx) = \Sigma_{az}^{\text{pas}}(\bx,z=0) + \Sigma^{\text{pol}}_{az}(\bx,z=0)
\end{equation}
represents the traction force per unit area that the cell exerts on the substrate \footnote{In averaging the force balance equation we have used the boundary condition $\Sigma_{az}^{\text{pas}}(z=h) \approx 0$. This indeed holds for the thermodynamic stress, which is given by $\Sigma_{az}^{\text{pas}}(z=h) = \kappa_{\icmp}\xi\phi_{\icmp}(\nabla_a\phi_{\icmp})/(2h) \sim \O(\xi/h)$, and thus is negligible in the limit $\xi \ll h$.}. Given that cell motility is generated through cell-substrate interaction, we assume that the magnitude of this stress is most dominant at the boundary in the shear components of the surface normal to the substrate [i.e., $\Sigma_{az}^{\text{pol}}(\bx,z=0) \gg h\oavg{\Sigma}_{ab}^{\text{pol}}(\bx)$]. Hence, we neglect the term $\oavg{\Sigma}_{ab}^{\text{pol}}$ in the averaged force balance equation and consider $\oavg{\Sigma}_{ab} \approx \oavg{\Sigma}_{ab}^{\text{pas}}$.

To make progress, we determine an expression for each side of \eqn\eqref{eqn:2D_traction_force_balance_single}, neglecting terms of $\O(\xi/h)$. For the left-hand side, we directly evaluate the $z$-average of the thermodynamic stress and then take the divergence. Using the fact that $\oavg{\phi_{\ocmp}^n}=1+\mathcal{O}(\xi/h)$ for any positive integer $n$ (\app\ref{app:z_average_phi}), we find
\begin{equation}  
  h\oavg{\Sigma}_{ab}^{\text{pas}} = -\delta_{ab}\Pi_{\icmp} - 2\kappa_{\icmp}\xi^2(\nabla_a\phi_{\icmp})(\nabla_b\phi_{\icmp}) \;,
\label{eqn:2D_passive_stressg_single}
\end{equation}
where $\Pi_{\icmp} = \mu_{_{\icmp}}\phi_{\icmp}-f^{\text{CH}}(\phi_{\icmp})+2\lambda_{\icmp}\delta A\phi_{\icmp}^2$ is a 2D osmotic pressure, with $\delta A = 1-A/A_0$ denoting the difference between the actual 2D cross-sectional cell area $A = \int\fd\bx\,\phi^2_{\icmp}$ and the preferred value $A_0$. Here, $\mu_{\icmp} = \delta  \F_\icmp / \delta\phi_{\perp}$ is a 2D chemical potential obtained from a 2D free energy
\begin{equation}
\begin{split}
  \F_\icmp =\;& \kappa_{\icmp}\int\fd\bx\,\left[f(\phi_{\icmp})+\xi^2(\vec{\nabla}_{\icmp}\phi_{\icmp})^2\right]
  \\
  &+\lambda_{\icmp}A_0\,\delta A^2 \;,
  \label{eq:2D_free_energy}
\end{split}
\end{equation}
where $\kappa_{\icmp} = h\kappa$ and $\lambda_{\icmp} = h\lambda$ (full calculations are shown in \app\ref{app:single}). The last term in \eqn\eqref{eq:2D_free_energy} is a soft constraint on the cell area. Under the approximation that the height $h$ of the cell is fixed, volume fluctuations directly determine area fluctuations, as observed in experiments on MDCK cell monolayers~\cite{zehnder2015cell}.
The assumption of a fixed cell thickness implies the vanishing of the $z$-component of the velocity at the cell interface. This in turn corresponds to assuming that the internal active forces giving rise to cell propulsion only contribute to the traction force that controls the in-plane cell motility. It would,  of course, be very interesting to also derive equations for the dynamics of the cell height, which can play an important role for instance in tissue folding or buckling \cite{salbreux2017mechanics,loewe2020shape,karzbrun2021human}. This is left for future work. By the same argument of how \eqn\eqref{eqn:3D_div_passive_stress_single} is related to \eqn\eqref{eqn:3D_passive_stress_single}, the divergence of this $z$-averaged stress is therefore
\begin{equation}
  \nabla_{b}\sigma_{ab} = h\nabla_b\oavg{\Sigma}_{ab} = - \phi_{\icmp}\nabla_a\mu_{\icmp} \;.
  \label{eqn:2D_force_density_single}
\end{equation}

For the traction force on the right-hand side of \eqn\eqref{eqn:2D_traction_force_balance_single}, we have $\Sigma_{az}^{\text{pas}}(\bx,z=0) \approx 0$ (see \app\ref{app:single}), and so $\rt_a \approx \Sigma_{az}^{\text{pol}}(\bx,z = 0)$. In line with previous theoretical studies on modeling cell-substrate interaction~\cite{loewe2020shape,banerjee2015propagating,banerjee2012contractile}, we write down this force phenomenologically as
\begin{equation}
  \bt(\bx) = \Gamma_{\icmp}\bv(\bx)-\bff(\bx) \;,
  \label{eqn:2D_traction_single}
\end{equation}
where $\Gamma_{\icmp}$ is a friction per unit area, which we assume to be constant, $\bff(\bx)$ is a propulsion force density, and $\bv(\bx)$ is the in-plane advection velocity field of the cell. We do not model the traction force at a subcellular level and thus do not assume a specific form for $\bff(\bx)$. However, previous approaches have done so via additional fields that describe actin dynamics within the cell~\cite{shao2010computational,ziebert2012model, moure2016computational}.

Finally, combining \eqns\eqref{eqn:2D_traction_force_balance_single}, \eqref{eqn:2D_force_density_single}, and \eqref{eqn:2D_traction_single}, we write the in-plane force balance statement as
\begin{equation}
  \Gamma_{\icmp}\bv(\bx) = \bff(\bx)-\phi_{\icmp}\vec{\nabla}_{\icmp}\mu_{\icmp} \;.
\end{equation}

\subsection{Particle-like equation of motion}

The in-plane force balance equation derived in the previous section allows one to solve for the cell velocity field, which can then be used for advecting the phase field of the cell over time. Generalizing this approach to multiple cells, whereby there are multiple velocity fields for individual cells, is challenging, as one needs to consider the boundary conditions between different fields at their interfaces, especially when these interfaces can evolve over time. To construct a framework that is simple to extend to the multicellular regime, we adopt a particle-like approach in describing the cell's kinematics, which is in line with the methods used in other phase-field modeling studies~\cite{mueller2019emergence,loewe2020solid,armengol2023epithelia}. Specifically, we define coarse-grained body vectors for the cell advection velocity $\bv^{\text{c}}$ and polarity $\bp^{\text{c}}$ with components
\begin{align}  
  \bv^{\text{c}} &\equiv \frac{\int\fd\bx\,\phi_{\icmp}\bv(\bx)}{\int\fd\bx\,\phi_{\icmp}}\;, \label{eqn:2D_CG_velocity_single} \\
  \bp^{\text{c}} &\equiv \frac{\int\fd\bx\,\phi_{\icmp}\bff(\bx)}{\Gamma_{\icmp} v_0 \int\fd\bx\,\phi_{\icmp}}\;,
  \label{eqn:2D_CG_polarity_single}
\end{align}
where $v_0$ is the magnitude of cell motility. Hence, the in-plane force balance equation becomes
\begin{equation}
  \Gamma_{\icmp}\bv^{\text{c}} = \Gamma_{\icmp} v_0 \bp^{\text{c}} - \frac{\int\fd\bx\,\phi_{\icmp}^2\vec{\nabla}_{\icmp}\mu_{\icmp}}{\int\fd\bx\,\phi_{\icmp}} \;.
  \label{eqn:2D_CG_force_balance}
\end{equation}
To incorporate noisy processes in a minimal way, we assume that the polarization vector
$\bp^{\text{c}} = (\cos\theta(t),\sin\theta(t))$ undergoes simple rotational diffusion with
\begin{equation}
  \fd\theta(t) = \sqrt{2D_r}\,\fd W(t) \;,
\label{eqn:dtheta}
\end{equation}
where $W(t)$ is a Wiener process and $D_r$ is the rotational diffusion constant. This form for the self-propulsion is the same as used previously in models of cells as deformable active Brownian particles~\cite{loewe2020solid,armengol2023epithelia}, with $\bv^{\text{c}} = v_0\bp^{\text{c}}$.

Although we have chosen the above form for the self-propulsion as a very simple model for activity, it is possible to incorporate other forms as well. For instance, it is possible to introduce velocity or shape alignment of the polarity by modifying \eqn\eqref{eqn:dtheta}, as considered in previous work (e.g.,~\cite{zhang2020active}), or by including the relevant cellular behavior in an evolution equation for the polarity vector itself~\cite{camley2017physical}.

Finally, with the cell velocity defined, one can determine the time evolution of the phase field $\phi$. Since this field is not conserved over time, its dynamics can be described by an advective-relaxational (Model A-like) equation of motion~\cite{hohenberg1977theory}: 
\begin{equation}
  \pd_t\phi_\icmp + \bv^{\text{c}} \cdot\vec{\nabla}_{\icmp} \phi_{\icmp} = -\frac{1}{\Gamma^{\phi}}\mu_{\icmp} \;,
  \label{eqn:2D_EOM_single}
\end{equation}
where $\Gamma^\phi$ is the inverse mobility that sets the characteristic relaxation time of the cell.

\section{Model for multiple cells}

We now extend the theory developed in the previous section to incorporate multiple cells, which are represented as individual phase fields. We begin by discussing changes in notation to accommodate multiple cells and modifications to the free energy and the force balance equation to account for intercellular interactions. We next perform the 3D-to-2D projection as in the single-cell case and derive a particle-like equation of motion for each cell.

\subsection{Notation, 3D free energy, and force balance}

The system which we consider now is that of a cell monolayer with $N$ cells. The monolayer has a uniform height $h$ (i.e., no cell-to-cell variation) and a lateral dimension $L$ in the $x$-$y$ plane. We denote with $\phi_i$  the phase field describing cell $i$ (with $i = 1,\dots, N$) and assume, as in the single-cell case, that it can be factorized into in-plane and out-of-plane fields. Since there is no variation in height, we describe all cells with the same height profile $\phi_z(z)$ [\eqn\eqref{eqn:phi_ocmp}] and write $\phi_i(\bX) = \phi_{i\icmp}(\bx)\,\phi_{\ocmp}(z), \forall i$.
Furthermore, we define
\begin{equation}
  \Phi \equiv \sum_{i=1}^N\phi_i = \phi_{\ocmp}\sum_{i=1}^N\phi_{i\icmp} \equiv \phi_z\,\Phi_{\icmp}\;,
\end{equation}
which describes the total tissue content (including water) at a specific point within the system. Note that summation involving cell indices $i$ is always explicitly written, unlike summation over indices for Cartesian components.

To take into account cell-cell interactions, we modify the free energy functional by incorporating a steric repulsion term to minimize cell overlap:
\begin{equation}
  \F^{\text{rep}} = \frac{\epsilon}{2}\sum_{i=1}^N\sum_{j\neq i}\int\fd\bX\,\phi_i^2\phi_j^2\;,
\end{equation}
with $\epsilon$ controlling the strength of the repulsion. The full free energy functional now reads
\begin{equation}
\begin{split}
  \F = \sum_{i=1}^N\bigg[ & \kappa\int\fd\bX\,\left[f(\phi_i) + \xi^2(\vec{\nabla}\phi_i)^2\right] + \lambda V_0\delta V_i^2\\ 
  & + \frac{\epsilon}{2} \sum_{j\neq i}\int\fd\bX\,\phi_i^2\phi_j^2\bigg] \;,
\end{split}
\end{equation}
where $f(\phi_i) = \phi_i^2\left(\phi_i-1\right)^2$ and $\delta V_i \equiv 1-V_i/V_0$, with $V_i = \int\fd\bX\,\phi_i^2$ being the volume of cell $i$.

Having multiple cells also changes the force balance equation. We consider three contributions to the total tissue stress:
\begin{equation}
  \Sigma_{\alpha\beta} = \Sigma_{\alpha\beta}^{\text{pas}} + \Sigma_{\alpha\beta}^{\text{pol}} + \Sigma_{\alpha\beta}^{\text{vis}} \;,
  \label{eqn:3D_stress}
\end{equation}
where $\Sigma_{\alpha\beta}^{\text{pas}}$ now denotes the passive thermodynamic stress on the cell monolayer, $\Sigma_{\alpha\beta}^{\text{pol}}$ the phenomenological stress associated with active processes that drive tissue propulsion, and $\Sigma_{\alpha\beta}^{\text{vis}}$ represents viscous dissipation arising from cell-cell friction. More precisely, the thermodynamic stress is defined by the relation
\begin{equation}
  \nabla_{\beta}\Sigma_{\alpha\beta}^{\text{pas}} \equiv -\sum_{i=1}^N \phi_i\nabla_{\alpha}\mu_i \;,
\end{equation}
and, with some algebra (\app\ref{app:passive_stress}), we obtain
\begin{equation}
\begin{split}
  \Sigma_{\alpha\beta}^{\text{pas}} = \sum_{i=1}^N\Bigg[& -\delta_{\alpha\beta}\Bigg(\Pi_i-\frac{\epsilon}{2}\sum_{j\neq i}\phi_i^2\phi_j^2\Bigg) \\
  &- 2\kappa\xi^2(\nabla_{\alpha}\phi_i)(\nabla_{\beta}\phi_i)\Bigg] \;,
  \label{eqn:3D_passive_stress}
\end{split}
\end{equation}
where $\Pi_i = \mu_i\phi_i-f^{\text{CH}}(\phi_i)+2\lambda\delta V_i\phi_i^2$ and $\mu_i = \delta\F/\delta\phi_i$ is the exchange chemical potential of cell $i$, given by
\begin{equation}
\begin{split}
  \mu_i =\;& 2\kappa\left[\phi_i(\phi_i-1)(2\phi_i-1)-\xi^2\nabla^2\phi_i\right] \\
  & - 4\lambda\delta V_i\phi_i + 2\epsilon\phi_i\sum_{j\neq i} \phi_j^2\;.
\end{split}
\end{equation}
The last term $\Sigma_{\alpha\beta}^{\text{vis}}$ in \eqn\eqref{eqn:3D_stress} accounts for intercellular friction by modeling it as a viscous-like stress in the tissue, given by 
\begin{equation}
  \Sigma_{\alpha\beta}^{\text{vis}} = \eta\left(\nabla_{\alpha}\rV_{\beta}+\nabla_{\beta}\rV_{\alpha}\right)+\left(\zeta-\frac{2}{3}\eta\right)\delta_{\alpha\beta}\nabla_{\gamma}\rV_{\gamma} \;,
\end{equation}
where $\eta$ and $\zeta$ are the shear and bulk viscosity, respectively \footnote{We do not explicitly write down a pressure term since the tissue flow is compressible -- i.e., mathematically, we do not impose a Lagrange multiplier to enforce incompressibility. Note also that the thermodynamic stress already contains an isotropic part that acts like a pressure.}. $\rV_{\alpha}$ are components of a coarse-grained tissue velocity or flow field, and we assume the in-plane components can factorize as 
\begin{equation}
  \rV_a(\bX) = \rV_{\icmp,a}(\bx)\,\mathcal{V}(z) = \rV_{\icmp,a}(\bx)\,\frac{z}{h^2}(2h-z)
\end{equation}
and $\rV_z = 0$ (i.e., the tissue motion in the $z$ direction is negligible compared to its lateral motion). The parabolic velocity profile $\mathcal{V}(z)$ takes into account that the cell is fixed to the substrate at $z = 0$ [i.e., no-slip boundary conditions (B.C.) with $\rV_a(\bx,0) = 0$] and has a free boundary at $z = h$ [i.e., slip B.C. with $\rV_a(\bx,h) = \rV_{\icmp,a}(\bx)$]. The in-plane profile $\bV_{\icmp}$ takes the form
\begin{equation}
  \bV_{\icmp}(\bx) \equiv \frac{1}{\Phi_{\icmp}}\sum_{i=1}^N\phi_{i\icmp}\bv_{i}^{\text{c}} \;,
  \label{eqn:2D_CG_velocity}
\end{equation}
where $\bv_{i}^{\text{c}}$ is the coarse-grained, body velocity vector of cell $i$ as introduced in the single-cell model [see \eqn\eqref{eqn:2D_CG_velocity_single}]. As shown later, this mathematical form of the stress and the tissue flow profile encapsulates the pairwise friction between cells of the form $\bv_{i}^{\text{c}}-\bv_{j}^{\text{c}}$ at the particle level.

\subsection{From 3D to 2D dynamics}

To formulate 2D in-plane equations of motion for the cells, we start from the in-plane components of the force balance equations ($\nabla_{\beta}\Sigma_{a\beta} = 0$) and average them over $z$, using the same boundary conditions as in the single-cell case [i.e., $\Sigma_{az}(\bx,z=0)\neq 0$ and $\Sigma_{az}(\bx,z=h) \approx 0$]. Similar to before, we find 
\begin{equation}
  \nabla_b\sigma_{ab} = h\nabla_{b}\left(\oavg{\Sigma}_{ab}^{\text{pas}} + \oavg{\Sigma}_{ab}^{\text{vis}}\right) = \rT_a(\bx) \;,
  \label{eqn:2D_force_balance}
\end{equation}
with the tissue traction force density $\rT_a(\bx)$ given by
\begin{equation}
  \rT_{a}(\bx) = \Sigma_{az}^{\text{pol}}(\bx,z=0) + \Sigma_{az}^{\text{vis}}(\bx,z=0) \;.
  \label{eqn:2D_traction}
\end{equation}
In writing down these equations, we have used the fact that $\Sigma_{az}^{\text{pas}}(\bx,z=0) \approx 0$ (see \app\ref{app:multi_average}) and made the assumption $\Sigma_{az}^{\text{pol}}(\bx,z=0) \gg h\oavg{\Sigma}_{ab}^{\text{pol}}(\bx)$, as done in the single-cell case.

To proceed further, we explicitly evaluate the $z$-averages of the stresses in \eqn\eqref{eqn:2D_force_balance} and the traction contributions in \eqn\eqref{eqn:2D_traction}. Similar to the single-cell case, the divergence of the averaged in-plane thermodynamic stress is
\begin{equation}
  \rF_{a}^{\text{pas}} \equiv h\nabla_b\oavg{\Sigma}_{ab}^{\text{pas}} = - \sum_{i=1}^N \phi_{i\icmp}\nabla_a\mu_{i\icmp} \;,
  \label{eqn:2D_passive_stress}
\end{equation}
where $\mu_{i\icmp}$ is the 2D exchange chemical potential for cell $i$ (see \app\ref{app:multi_average} for details). For the divergence of the averaged in-plane viscous stress, we obtain
\begin{equation}
  \rF_{a}^{\text{vis}} \equiv h\nabla_b\oavg{\Sigma}_{ab}^{\text{vis}}
  = \eta_{\icmp}\nabla_b\nabla_b\rV_{\icmp,a} + \nu_{\icmp}\nabla_a\nabla_b\rV_{\icmp,b} \;,
  \label{eqn:2D_viscous_stress}
\end{equation}
where $\nu_{\icmp}=\zeta_{\icmp}+\eta_{\icmp}/3$, $\eta_{\icmp}=2\eta h/3$, and $\zeta_{\icmp}=2\zeta h/3$.

For the traction, we again treat the term originated from active cell processes driving self-propulsion phenomenologically and write
\begin{equation}
  \Sigma_{az}^{\text{pol}}(\bx,z=0) = \Gamma_{\icmp}^{\text{sub}}\rV_{\icmp,a}(\bx) - \rF_{a}^{\text{pol}}(\bx) \;,
\end{equation}
where $\Gamma_{\icmp}^{\text{sub}}$ is a friction constant and $\rF_{a}^{\text{pol}}$ are components of a coarse-grained tissue force field constructed from the propulsion of individual cells (see definition below). An additional contribution to the traction arises from the viscous stress, which has the same form as a friction (\app\ref{app:multi_average}):
\begin{equation}
  \Sigma_{az}^{\text{vis}}(\bx,z=0) = \frac{3\eta_{\icmp}}{h^2}\rV_{\icmp,a}(\bx) \equiv \Gamma_{\icmp}^{\text{vis}}\,\rV_{\icmp,a}(\bx) \;.
  \label{eq:viscous_traction}
\end{equation}
Combining the two contributions to the traction, we find
\begin{equation}
  \bT(\bx) = -\bF^{\text{sub}} - \bF^{\text{pol}} = \Gamma_{\icmp}\left[\bV_{\icmp}(\bx) - v_0 \bP_{\icmp}(\bx)\right] \;,
\end{equation}
where we have defined $\bF^{\text{sub}} \equiv -\Gamma_{\icmp}\bV_{\icmp}(\bx)$,
with $\Gamma_{\icmp} \equiv \Gamma_{\icmp}^{\text{sub}} + \Gamma_{\icmp}^{\text{vis}}$, and that $\bP_{\icmp}(\bx)$ is the tissue polarization field given by
\begin{equation}
  \bP_{\icmp}(\bx) = \frac{\bF^{\text{pol}}}{\Gamma_{\icmp} v_0} \equiv \frac{1}{\Phi_{\icmp}}\sum_{i=1}^N\phi_{i\icmp}\bp_{i}^{\text{c}} \;,
\label{eq:2D_CG_polarity}
\end{equation}
with $\bp_{i}^{\text{c}}$ being the coarse-grained polarity vector of cell $i$ [\eqn\eqref{eqn:2D_CG_polarity_single}].
Finally, putting together all the results in this section and neglecting terms of $\O(\xi/h)$, we obtain the following multicellular force balance equation for the tissue
\begin{equation}
  \bF^{\text{pas}} + \bF^{\text{sub}} + \bF^{\text{pol}} + \bF^{\text{vis}} = \vec{0} \;.
  \label{eqn:2D_tissue_force_balance}
\end{equation}

\subsection{Solving for cell velocities and the equations of motion for individual phase fields}

The tissue force balance equation that we just derived encodes the dynamics of all interactions between cells. We now utilize this equation to extract the body velocities $\bv_{i}^{\text{c}}$ of individual cells, which are needed in the equations of motion for updating the phase fields $\phi_i$, similar to the single-cell case [see \eqn\eqref{eqn:2D_EOM_single}]. Because the tissue velocity field $\bV_{\icmp}$, by construction, linearly interpolates $\bv_{i}^{\text{c}}$ [\eqn\eqref{eqn:2D_CG_velocity}], one can employ standard linear algebra techniques to solve for these velocities. 

To this end, we first observe that \eqn\eqref{eqn:2D_tissue_force_balance} is written at a continuum level for each position $\bx$, and that it is more convenient to work with this equation at the ``particle'' or cellular level. This can be done by projecting this equation onto each cell (say cell $i$), whereby we perform the operation
\begin{equation}
  \bff_{i}^{\text{Y}} = \int\fd\bx\,\phi_{i\icmp}\,\bF^{\text{Y}} \;,
\end{equation}
on each force term $\text{Y}$ and write the force balance equation for the cell as 
\begin{equation}
  \bff_{i}^{\text{pas}} + \bff_{i}^{\text{sub}} + \bff_{i}^{\text{pol}} + \bff_{i}^{\text{vis}} = \vec{0} \;.
  \label{eqn:2D_force_balance_cell}
\end{equation}
The first three force terms can be expressed as
\begin{align}
  \bff_{i}^{\text{pas}} &= -\sum_{j=1}^N \int\fd\bx\,\phi_{i\icmp}\phi_{j\icmp}\vec{\nabla}_{\icmp}\mu_{j\icmp} \;, \\
  \bff_{i}^{\text{sub}} &= -\Gamma_{\icmp}\sum_{j=1}^N O_{ij}\bv_{j}^{\text{c}} \;, \\
  \bff_{i}^{\text{pol}} &= \Gamma_{\icmp}v_0\sum_{j=1}^N O_{ij}\bp_{j}^{\text{c}} \;,
\end{align}
where we defined the degree of overlap between cells $i$ and $j$ to be
\begin{equation}
  O_{ij} \equiv \int\fd\bx\,\frac{\phi_{i\icmp}\phi_{j\icmp}}{\Phi_\icmp} \;.
\end{equation}
For the viscous force term, we find 
\begin{equation}
  \bff_{i}^{\text{vis}} = \sum_{j=1}^N\int\fd\bx\,\vec{M}_{ij}\cdot\left(\bV_{\icmp}-\bv_{j}^{\text{c}}\right) \;,
  \label{eq:viscous_force_projection}
\end{equation}
where 
\begin{equation}
  \vec{M}_{ij} = \frac{1}{\Phi_{\icmp}}\left[\eta_{\icmp}\trace\left(\vec{D}_{ij}\right)\vec{1} +\nu_{\icmp}\vec{D}_{ij}\right]
  \label{eq:viscous_interface_matrix}    
\end{equation}
with $\vec{D}_{ij} = (\vec{\nabla}_{\icmp}\phi_{i\icmp})(\vec{\nabla}_{\icmp}\phi_{j\icmp})^T$. Here, $\vec{M}_{ij}$ is a tensor associated with the extent of interfacial overlap between cells $i$ and $j$. Full details of this calculation can be found in \app\ref{app:project}. Indeed, by making the approximation that three-body overlaps are rare (i.e., products involving three different phase fields and/or their gradients are negligible), one can further show that 
\begin{equation}
  \bff_{i}^{\text{vis}} \approx \sum_{j\neq i}\int\fd\bx\,\widetilde{\vec{M}}_{ij}\cdot\left(\bv_{i}^\text{c}-\bv_{j}^{\text{c}}\right) \;,
\end{equation}
where $\widetilde{\vec{M}}_{ij}$ contains terms similar to $\vec{M}_{ij}$ (see \app\ref{app:project}). Hence, the viscous stress that we formulated at the tissue level allows us to recover pairwise friction between cells at the cellular level.

To make further progress, we note that only $\bff_{i}^{\text{sub}}$ and $\bff_{i}^{\text{vis}}$ have terms that are linear in $\bv_{i}^{\text{c}}$, while other force terms do not. This motivates us to write the cellular level force balance equation [\eqn\eqref{eqn:2D_force_balance_cell}] as a matrix equation with coupled advection velocities. Using the identity
\begin{equation}
  \sum_{j=1}^N(\vec{\nabla}_{\icmp}\phi_{j\icmp})\cdot\bV_{\icmp} = \sum_{j=1}^N\frac{\phi_{j\icmp}}{\Phi_{\icmp}}(\vec{\nabla}_{\icmp}\Phi_{\icmp})\cdot\bv_{j}^{\text{c}} \;,
\end{equation}
we rewrite the viscous force on cell $i$ as
\begin{equation}
  \bff_{i}^{\text{vis}} = -\sum_{j=1}^N\left[\eta_{\icmp}\trace\left(\vec{K}_{ij}\right)\vec{1} + \nu_{\icmp}\vec{K}_{ij}\right]\cdot\bv_{j}^{\text{c}} \;,
\end{equation}
where $\vec{K}_{ij}$ has components
\begin{equation}
  K_{ij,ab} = \int\fd\bx\,\frac{\nabla_a\phi_{i\icmp}}{\Phi_\icmp}\left[\nabla_b\phi_{j\icmp}-\frac{\phi_{j\icmp}(\nabla_b\Phi_{\icmp})}{\Phi_{\icmp}}\right]\;.
\end{equation}
By defining
\begin{equation}
  \vec{A}_{ij} \equiv \left[\eta_{\icmp}\trace\left(\vec{K}_{ij}\right) + \Gamma_{\icmp}O_{ij}\right]\bm{1} + \nu_{\icmp}\vec{K}_{ij} 
\end{equation}
and $\bff_i^{\text{c}} \equiv \bff_i^{\text{pas}} + \bff_i^{\text{pol}}$, which does not depend on $\bv_i^{\text{c}}$, we can now cast the force balance equation in the following matrix form
\begin{equation}
  \begin{pmatrix}
    \vec{A}_{11} & \vec{A}_{12} & \cdots & \vec{A}_{1N} \\
    \vec{A}_{21} & \vec{A}_{22} & \cdots & \vec{A}_{2N} \\
    \vdots & \vdots & \ddots & \vdots & \\
    \vec{A}_{N1} & \vec{A}_{N2} & \cdots & \vec{A}_{NN} \\
  \end{pmatrix} 
  \begin{pmatrix}
    \vec{\text{\textbf{v}}}^{\text{c}}_{1} \\ \vec{\text{\textbf{v}}}^{\text{c}}_{2} \\ \vdots \\ \vec{\text{\textbf{v}}}^{\text{c}}_{N}
  \end{pmatrix} = 
  \begin{pmatrix}
    \bff_1^{\text{c}} \\ \bff_2^{\text{c}} \\ \vdots \\ \bff_N^{\text{c}}
  \end{pmatrix}\;.
  \label{eqn:cell_advection_matrix}
\end{equation}
As a result, the body velocity of each cell $\bv_i^{\text{c}}$ can be found by inverting this matrix equation. This velocity is then used to update the phase field $\phi_{i\icmp}$, which obeys the following equation of motion:
\begin{equation}
  \pd_t\phi_{i\icmp} + \bv_i^{\text{c}} \cdot\vec{\nabla}_{\icmp}\phi_{i\icmp} = -\frac{1}{\Gamma^{\phi}}\mu_{i\icmp}\;.
  \label{eqn:2D_EOM}
\end{equation}
Solving \eqns\eqref{eqn:cell_advection_matrix} and \eqref{eqn:2D_EOM} iteratively by numerical means allows us to simulate the dynamics of the cell monolayer over time using a particle-like approach.

\begin{figure}
  \centering
  \includegraphics[width=\linewidth]{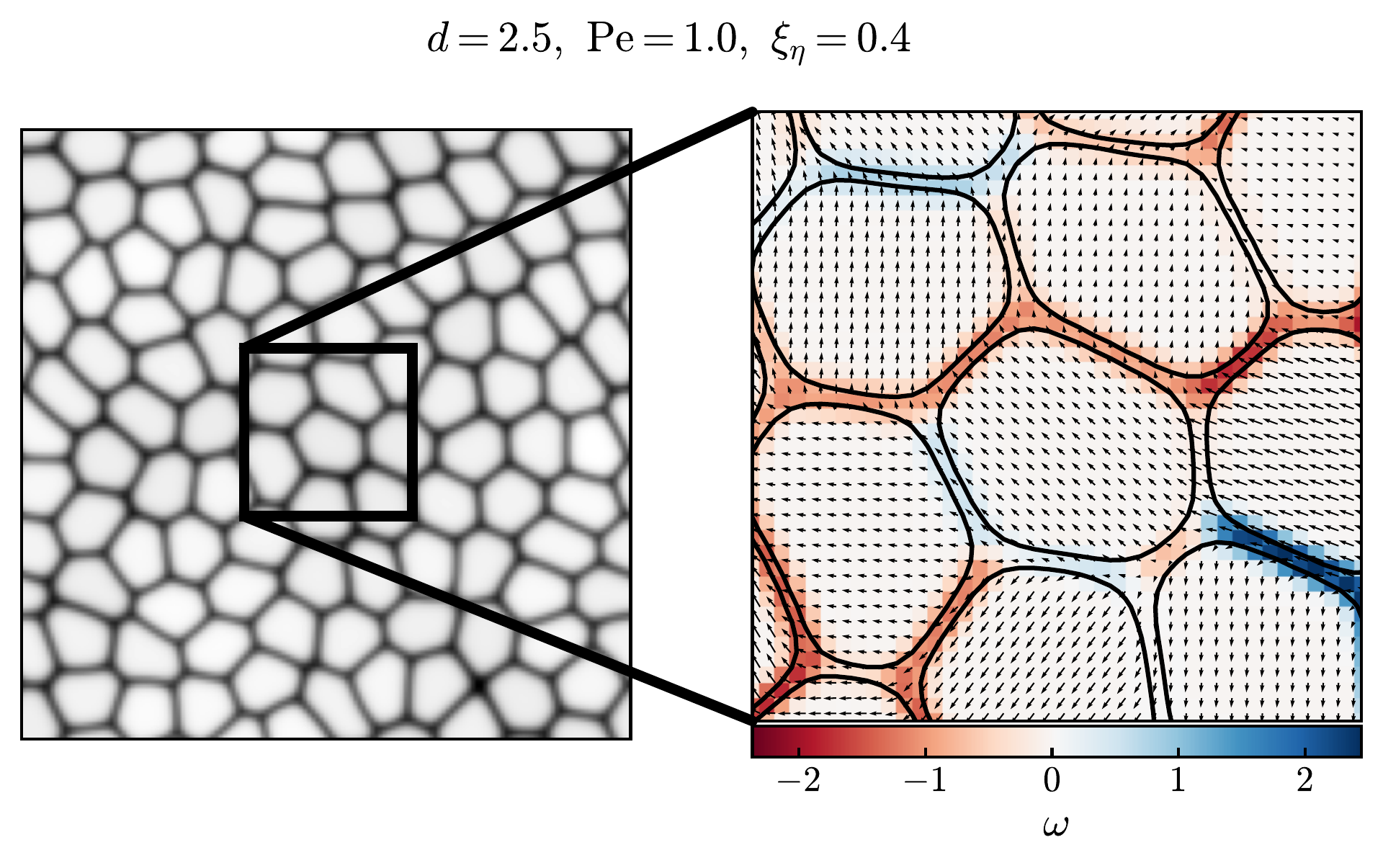}
  \caption{\textit{Left}: A representative snapshot of the total field $\Phi_\perp$ for a system with $d = 2.5$, $\text{Pe} = 1.0$, and $\xi_{\eta} = 0.4$. \textit{Right}: A zoomed-in snapshot showing the cell contours (black lines), the tissue velocity field $\bV_{\icmp}(\bx)$, as defined in \eqn\eqref{eqn:2D_CG_velocity} (black arrows), and the color indicating its vorticity magnitude. While the velocity field is constant within a cell, there can be spatial gradients at the interfaces between cells, as shown by the non-zero vorticity in those regions.}
  \label{fig:tvf}
\end{figure}

To illustrate how the flow field constructed in \eqn\eqref{eqn:2D_CG_velocity} can model intercellular friction, we show a snapshot of the full system in \fig\ref{fig:tvf} (\textit{left}) alongside the velocity field (\textit{right}). The velocity field is constant within cell contours, but there are shear flows in between cells, which control the cell-cell frictional forces. In \eqn\eqref{eq:viscous_force_projection}, we determine the net effect of the forces resulting from these flows on each cell's velocity.

\begin{figure*}[t]
  \centering
  \includegraphics[width=\linewidth]{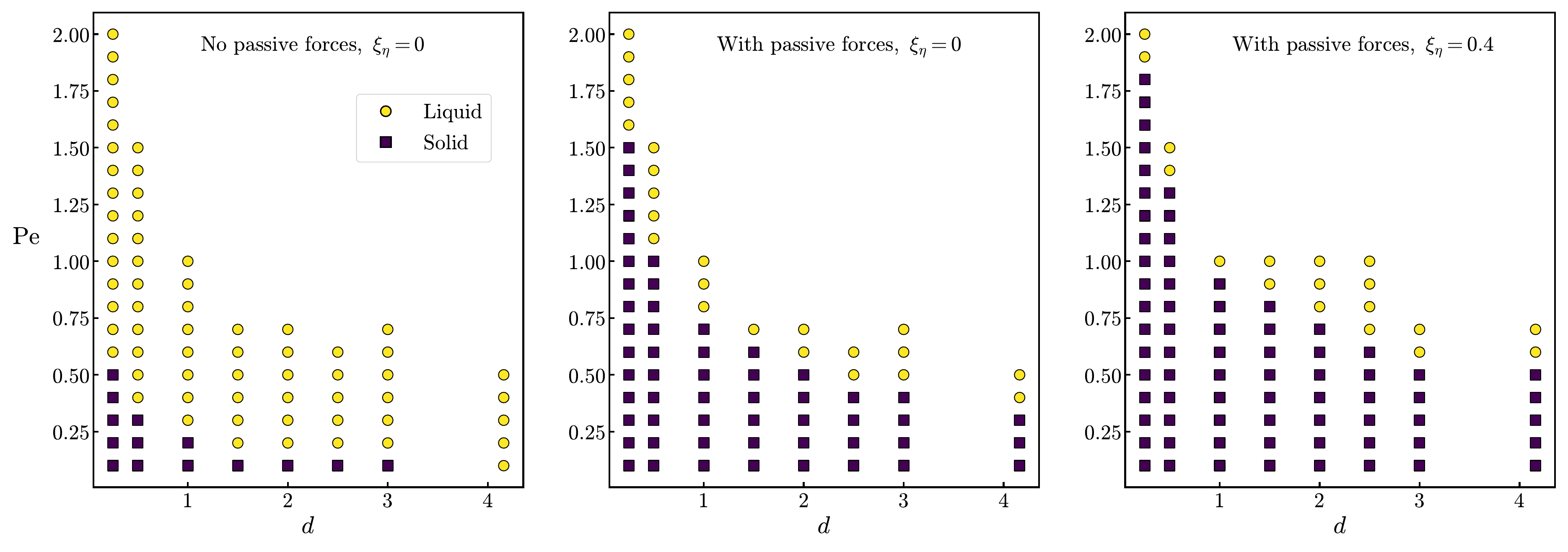}
  \caption{Solid-liquid phase diagrams of a 2D cell monolayer in the plane of deformability $d$ and P\'eclet number $\text{Pe}$. Yellow circles denote the liquid phase, and purple squares denote the solid phase. The left frame corresponds to the case where both passive thermodynamic forces and cell-cell friction are neglected in the force balance equation for the cell velocity. The phase diagram shown in the middle frame is obtained by including passive forces, but not cell-cell friction. Both forces are included in the right frame.}
  \label{fig:pd}
\end{figure*}

\pagebreak
\section{Effect of the passive stress on the solid-liquid transition}

Previous work employing phase-field models to describe cell monolayers has either included or neglected passive thermodynamic forces in the force balance equation for each cell. To quantify the effect of these forces on the rheological state of the monolayer, we simulate systems of $100$ cells at various cell-edge tensions $\gamma_{\icmp} = \kappa_{\icmp}\xi/3$ and self-propulsion velocities $v_0$ (see \app\ref{app:sims} for simulation details and numerical schemes used). We characterize the relative importance of the cell-edge tension to steric repulsion via the deformability parameter $d = \epsilon_{\icmp} \xi R / (12 \gamma_{\icmp} R)$, a dimensionless number which measures the extent to which cells in contact change their shape or overlap. Previous work has shown that tuning cell deformability yields qualitative changes in the melting and rheology of the monolayer~\cite{loewe2020solid, hopkins2022local}. The cells' motility is quantified by the P{\'e}clet number $\text{Pe}=v_0/(R D_r)$, which is the ratio of the cell's persistence length to its size. Finally, we also examine the role of cell-cell friction controlled by the viscosity $\eta_{\icmp}$, which sets the length scale $\xi_{\eta}=\sqrt{\eta_{\icmp}/\Gamma_{\icmp}}$ over which the velocities between different cells are correlated.

\fig\ref{fig:pd} shows the solid-liquid phase diagrams for the system as determined by the effective diffusivity $D_{\text{eff}} = \lim_{t\to\infty}\text{MSD}(t)/(4 D_0 t)$, where $D_0=v_0^2/(2 D_r)$ is the diffusivity of an isolated cell, and where we choose a cutoff of $D_{\text{eff}} = 0.002$ to differentiate the liquid and solid states. It is evident that while the three phase diagrams show the same qualitative behavior, both passive thermodynamic forces and cell-cell friction tend to solidify the monolayer, shifting the melting line to larger deformability and motility.

\section{A comparison between adhesion and intercellular friction}
\label{sect:adhesion}  
\begin{figure}
  \centering
  \includegraphics[width=\linewidth]{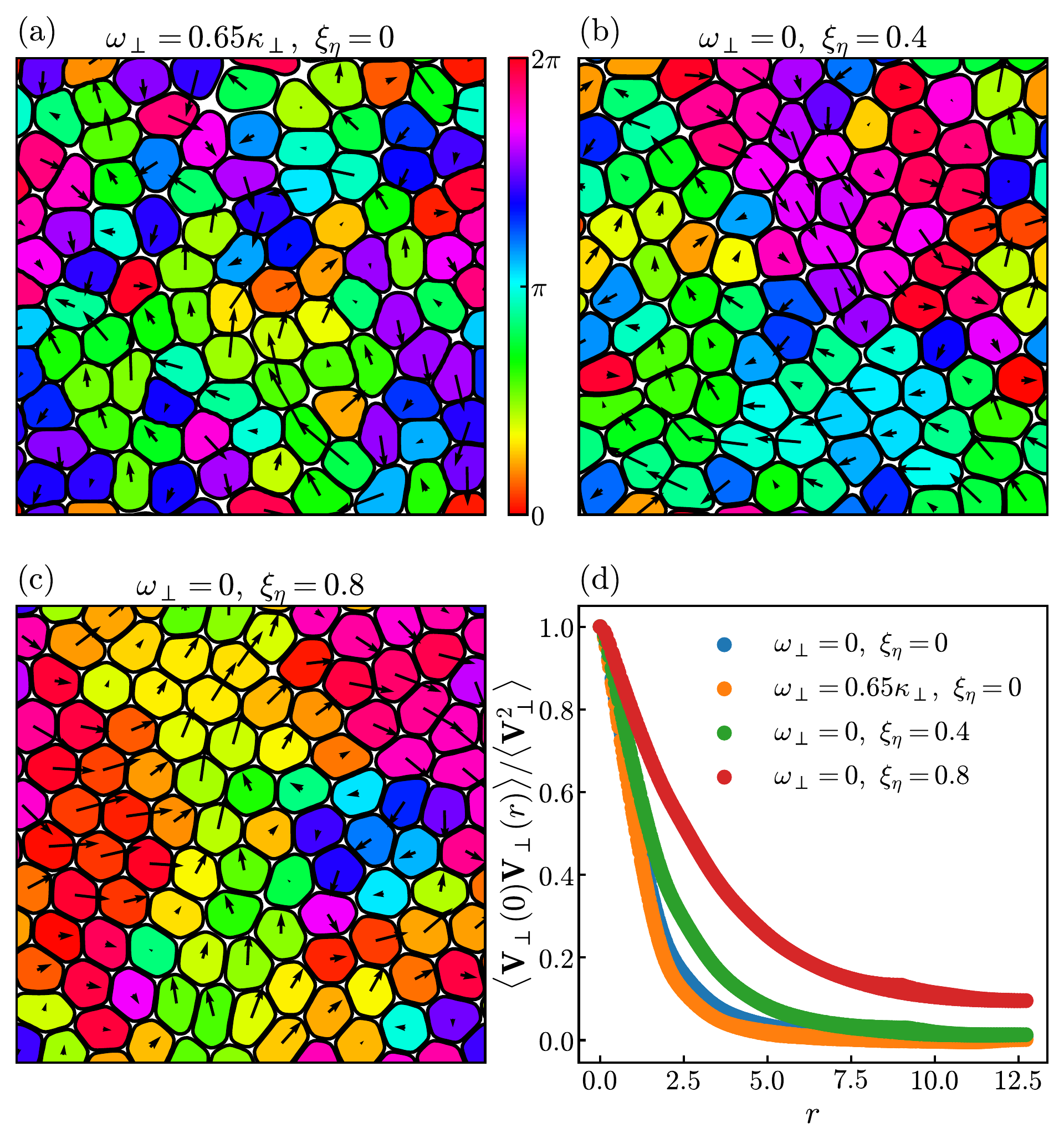}
  \caption{Intercellular friction generates stronger spatial velocity correlations than an energetic adhesive interaction. (a)--(c) Representative snapshots of a system with $d = 2.5$ and $\text{Pe} = 1.0$, where the cells are colored by the angle of their center-of-mass velocity. In (a), only adhesion ($\omega_\icmp/\kappa_\icmp = 0.65)$ is present, while (b) and (c) show that spatial velocities correlations increase with increasing $\xi_\eta$. (d) Corresponding spatial correlation functions of the tissue velocity field $\bV_{\icmp}(\bx)$ defined in \eqn\eqref{eqn:2D_CG_velocity}, averaged over time. A system without friction and adhesion (blue dots) has velocity correlations nearly identical to one with only adhesion (orange dots).}
  \label{fig:vel_cor}
\end{figure}

Within the literature on phase field models, various adhesive interactions between cells have been incorporated into the free energy to model cadherin bonds between cells (e.g.,~\cite{nonomura2012study,najem2016phase,monfared2023mechanical,kuang2023morphosim,zhang2020active,zadeh2022picking}). One significant difference between these adhesive interactions and the intercellular friction we have derived here is that the latter depends on the relative cell velocities, while energetic adhesive interactions only depend on the instantaneous configuration. Experiments on epithelial monolayers have shown that cell-cell adhesion affects the correlation of cell velocities~\cite{czirok2013collective}. While a complete study on the full effect of cell-cell adhesion and intercellular friction on the velocity correlations in tissue is beyond the scope of this work, we provide below a comparison of the effect of the two interactions on the velocity correlations of the cells.

To model adhesion, we introduce an additional contribution to the free energy, given by
\begin{equation}
  \F^{\text{adh}} = \frac{\omega\xi^2}{2}\sum_{i=1}^N\sum_{j\neq i}\int\fd\bX\,\left(\vec{\nabla} \phi_i^2\right) \cdot \left(\vec{\nabla}\phi_j^2\right)\;,
\end{equation}
with $\omega > 0$ controlling the adhesive coupling strength. Consistent with our definition of the cell volume and our formulation of steric repulsion, we use the square of the field in the free energy rather than the field itself, which results in a chemical potential for the $i$th cell that always depends on its field $\phi_i$, thus giving a more physical evolution equation for the phase field. Using an averaging procedure similar to that in \app \ref{app:multi_average}, we find the contribution in 2D to be the same as that would have come from a 2D adhesion energy term of the same form with the in-plane fields $\phi_{i\icmp}$ and a coefficient $\omega_\icmp = h\omega$.

The adhesion energy gives an attractive interaction between different cells, but it only impacts the velocity indirectly through the passive forces in the force balance equation. In contrast, the intercellular friction arising from viscous stresses at the tissue level directly couples the velocities of neighboring cells. We show in \fig\ref{fig:vel_cor} the effect of these two types of coupling on velocity correlations for a system with $d = 2.5$ and $\text{Pe} = 1.0$. In the snapshots, the cells are colored by the orientation of their center-of-mass velocity. It is evident that intercellular friction [\figs\ref{fig:vel_cor}(b) and (c)] enhances the correlated movement of neighboring cells relative to adhesion [\fig\ref{fig:vel_cor}(a)]. This difference can be quantified by the spatial correlation function of the tissue velocity shown in \fig\ref{fig:vel_cor}(d), which decays sharply when only adhesion is present. Energetic adhesive interactions essentially yield the same spatial velocity correlations as a system without friction and adhesion. Yet, velocity correlations grow significantly with increasing intercellular friction $\xi_\eta$. We expect that both energetic adhesive interactions and intercellular friction are controlled by cadherins. Both may be necessary for a proper description of cell-cell couplings that captures cell shapes and velocity correlations, but we defer a detailed study of this to future work.

\section{Summary and Discussion}

2D phase field models have often been used in place of 3D ones due to the large computational cost of numerically solving the equations for each cell's field. Here, we have shown that, under certain assumptions, the connection between 3D and 2D phase field models for cells on a substrate can be made explicit by averaging over the out-of-plane direction. In doing so, we arrive at a model of cells on a substrate as deformable particles, where their shape is described by a 2D field, which is advected by a body velocity vector determined from force balance. A similar approach could be used to derive equations for phase field models which include additional interactions, such as adhesion or active stresses from sources other than traction (e.g., stresses transmitted via adherens junctions).

We have also considered the effect of intercellular friction on the solid-liquid transition of the monolayer. Other work~\cite{peyret2019sustained} has phenomenologically incorporated viscous interactions as a function of differences in cell polarity, but here we show that the effect of friction between cells can be derived from a viscous interaction of the cell velocities. We then show that these viscous forces result in an increase in the required motility for the monolayer to melt, at low and high cell deformabilities. We also show that the viscous forces introduce spatial correlations in the velocities, which are not captured by an energetic adhesive interaction between cells. This model will enable future work to further elucidate the role of intercellular friction in cell monolayers.

\begin{acknowledgments}
We thank Amin Doostmohammadi for valuable discussions. The work by A.H. and M.C.M. was supported by the National Science Foundation Grant No. DMR-2041459. This research has received funding (B.L.) from the European Research Council under the European Union’s Horizon 2020 research and innovation programme (Grant Agreement No. 851196). Use was made of computational facilities purchased with funds from the National Science Foundation (CNS-1725797) and administered by the Center for Scientific Computing (CSC). The CSC is supported by the California NanoSystems Institute and the Materials Research Science and Engineering Center (MRSEC; NSF DMR 2308708) at UC Santa Barbara. For the purpose of open access, the author has applied a Creative Commons Attribution (CC BY) licence to any Author Accepted Manuscript version arising from this submission.\vspace{5pt}
\end{acknowledgments}

M.C. and A.H. contributed equally to this work.

\appendix

\section{Deriving the passive\\ thermodynamic stress}
\label{app:passive_stress}

In this appendix, we first review the derivation of the constitutive relation for the passive thermodynamic stress $\Sigma_{\alpha\beta}^{\text{pas}}$ for the single-cell case (see also~\cite{cates2018theories}), before extending the theory to the case with multiple cells. Recall that the single-cell free energy is
\begin{equation}
  \F = \F^{\text{CH}} + \F^V = \int\fd\bX\,f^{\text{CH}}(\phi,\vec{\nabla}\phi) + \lambda V_0\delta V^2 \;,
\end{equation}
where the Cahn-Hilliard free energy density is
\begin{equation}
  f^{\text{CH}} = \kappa\left[f(\phi)+\xi^2\left(\vec{\nabla}\phi\right)^2\right] 
\end{equation}
with
\begin{equation}
  f(\phi) = \phi^2(\phi-1)^2 \;,
\end{equation}
and that the passive force density arising from deformations that cause the cell to deviate from the ideal profile, as set by the free energy, is
\begin{equation}
  \nabla_{\beta}\Sigma_{\alpha\beta}^{\text{pas}} = -\phi\nabla_{\alpha}\mu = -\nabla_{\alpha}(\mu\phi) + \mu\nabla_{\alpha}\phi \;.
\end{equation}
Here, the exchange chemical potential $\mu$ is given by
\begin{equation}
  \mu = \frac{\delta\F}{\delta\phi} = \frac{\pd f^{\text{CH}}}{\pd\phi} - \nabla_{\beta}\frac{\pd f^{\text{CH}}}{\pd\nabla_{\beta}\phi} + \mu^V \;,
\end{equation}
with $\mu^V = \delta\F^V/\delta\phi$. Using the fact that
\begin{equation}
  \nabla_{\alpha}f^{\text{CH}} = \frac{\pd f^{\text{CH}}}{\pd\phi}\nabla_{\alpha}\phi + \frac{\pd f^{\text{CH}}}{\pd\nabla_{\beta}\phi}\nabla_{\alpha}\nabla_{\beta}\phi \;,
\end{equation}
we obtain
\begin{equation}
\begin{split}
  \mu\nabla_{\alpha}\phi =\;& \nabla_{\alpha}f^{\text{CH}}-\nabla_{\beta}\left(\frac{\pd f^{\text{CH}}}{\pd\nabla_{\beta}\phi}\nabla_{\alpha}\phi\right) + \mu^V\nabla_{\alpha}\phi \\
  =\;& \nabla_{\beta}\big[-\delta_{\alpha\beta}\left(2\lambda\delta V\phi^2-f^{\text{CH}}\right) \\
  & \quad\;\;\; - 2\kappa\xi^2(\nabla_{\alpha}\phi)(\nabla_{\beta}\phi)\big] \;,
\end{split}
\end{equation}
and so
\begin{equation}
  \nabla_{\beta}\Sigma_{\alpha\beta}^{\text{pas}} = \nabla_{\beta}\big[-\delta_{\alpha\beta}\Pi-2\kappa\xi^2(\nabla_{\alpha}\phi)(\nabla_{\beta}\phi)\big] \;,
\end{equation}
where $\Pi = \mu\phi-f^{\text{CH}}+2\lambda\delta V\phi^2$. As a result, we identify that the passive thermodynamic stress is given by
\begin{equation}
  \Sigma_{\alpha\beta}^{\text{pas}} = -\delta_{\alpha\beta}\Pi - 2\kappa\xi^2(\nabla_{\alpha}\phi)(\nabla_{\beta}\phi) \;,
\end{equation}
 as stated in \eqn\eqref{eqn:3D_passive_stress_single} in the main text.

Extending this theory to multiple phase fields requires some careful analysis. Na\"{i}vely, one might think that the passive stress can be constructed by adding the corresponding stress contributions from individual cells. This is valid as long as there are no interactions between cells, which is not the case here, as there is steric repulsion between them. As we shall see, it is not possible, in general, to write down the passive force density on each cell element as the divergence of a cellular stress tensor due to the intercellular terms. We begin by writing down the free energy for multiple cells, which can be expressed in the following form:
\begin{equation}
\begin{split}
  \F =\;& \sum_{i=1}^N\Bigg( \int\fd\bX\,f_i^{\text{CH}} + \lambda V_0\delta V_i^2\Bigg) \\
  &+ \frac{1}{2}\sum_{i=1}^N\sum_{j\neq i}\int\fd\bX\,f_{ij}^{\text{int}} \;,
\end{split}
\end{equation}
where we have assumed all interactions are pairwise and denoted
\begin{align}
  f_i^{\text{CH}} &\equiv f^{\text{CH}}(\phi_i,\vec{\nabla}\phi_i) \;, \\
  f_{ij}^{\text{int}} &\equiv f^{\text{int}}(\phi_i,\phi_j,\vec{\nabla}\phi_i,\vec{\nabla}\phi_j) \;.
\end{align}
We also assume that these interactions are symmetric
\begin{equation}
  f_{ij}^{\text{int}} = f_{ji}^{\text{int}} \;,
\end{equation}
and this is indeed the case within our model, since $f_{ij}^{\text{int}} = \epsilon\phi_i^2\phi_j^2$ [with the additional contribution of $\omega\xi^2(\vec{\nabla}\phi_i^2)\cdot(\vec{\nabla}\phi_j^2)$ when there is cell-cell adhesion (see Section~\ref{sect:adhesion})]. From the free energy, we find the exchange chemical potential for cell $i$ to be
\begin{equation}
\begin{split}
  \mu_i = \frac{\delta\F}{\delta\phi_i} =\;& \frac{\pd f_i^{\text{CH}}}{\pd\phi_i}-\nabla_{\beta}\frac{\pd f_i^{\text{CH}}}{\pd\nabla_{\beta}\phi_i} + \mu_i^V \\
  & + \frac{1}{2}\sum_{j=1}^N\sum_{k\neq j}\left(\frac{\pd f_{jk}^{\text{int}}}{\pd\phi_i} - \nabla_{\beta}\frac{\pd f_{jk}^{\text{int}}}{\pd\nabla_{\beta}\phi_i}\right) \;.
\end{split}
\end{equation}
Here, the summed terms arise from pairwise interactions between cells. Since they are only non-zero when $j = i$ or $k = i$ and that $f_{jk}^{\text{int}}$ is symmetric, we can write
\begin{equation}
\begin{split}
  & \frac{1}{2}\sum_{j=1}^N\sum_{k\neq j}\left(\frac{\pd f_{jk}^{\text{int}}}{\pd\phi_i} - \nabla_{\beta}\frac{\pd f_{jk}^{\text{int}}}{\pd\nabla_{\beta}\phi_i}\right) \\
  =\;& \sum_{j\neq i}\left(\frac{\pd f_{ij}^{\text{int}}}{\pd\phi_i}-\nabla_{\beta}\frac{\pd f_{ij}^{\text{int}}}{\pd\nabla_{\beta}\phi_i}\right) \;.
\end{split}
\end{equation}
We now proceed with the same approach as the single-cell case to determine the constitutive relation for the thermodynamic stress. Let
\begin{equation}
  \rF_{i,\alpha}^{\text{pas}} \equiv -\phi_i\nabla_{\alpha}\mu_i = -\nabla_{\alpha}(\mu_i\phi_i) + \mu_i\nabla_{\alpha}\phi_i
\end{equation}
be the passive thermodynamic force density acting on cell element $i$. Using the relation
\begin{equation}
  \nabla_{\alpha}f_i^{\text{CH}} = \frac{\pd f_i^{\text{CH}}}{\pd\phi_i}\nabla_{\alpha}\phi_i + \frac{\pd f_i^{\text{CH}}}{\pd\nabla_{\beta}\phi_i}\nabla_{\alpha}\nabla_{\beta}\phi_i \;,
\end{equation}
we can write
\begin{equation}
\begin{split}
  \mu_i\nabla_{\alpha}\phi_i =\;& \nabla_{\alpha}f_i^{\text{CH}}-\nabla_{\beta}\left(\frac{\pd f_i^{\text{CH}}}{\pd\nabla_{\beta}\phi_i}\nabla_{\alpha}\phi_i\right) + \mu_i^V\nabla_{\alpha}\phi_i \\
  & + \sum_{j\neq i}\left(\frac{\pd f_{ij}^{\text{int}}}{\pd\phi_i}-\nabla_{\beta}\frac{\pd f_{ij}^{\text{int}}}{\pd\nabla_{\beta}\phi_i}\right)\nabla_{\alpha}\phi_i \;,
\end{split}
\end{equation}
and therefore
\begin{equation}
\begin{split}
  \rF_{i,\alpha}^{\text{pas}} =\;& \nabla_{\beta}\left(-\delta_{\alpha\beta}\Pi_i - \frac{\pd f_i^{\text{CH}}}{\pd\nabla_{\beta}\phi_i}\nabla_{\alpha}\phi_i\right) \\
  & + \sum_{j\neq i}\left(\frac{\pd f_{ij}^{\text{int}}}{\pd\phi_i}-\nabla_{\beta}\frac{\pd f_{ij}^{\text{int}}}{\pd\nabla_{\beta}\phi_i}\right)\nabla_{\alpha}\phi_i \;,
\end{split}
\end{equation}
where $\Pi_i = \mu_i\phi_i-f_i^{\text{CH}}+2\lambda\delta V_i\phi_i^2$. Clearly, the presence of the interaction terms makes it not possible to write $\rF_{i,\alpha}^{\text{pas}}$ as the divergence of a stress tensor. Nevertheless, we now show that the total thermodynamic force density, from summing $\rF_{i,\alpha}^{\text{pas}}$ over $i$, can be expressed as the divergence of a stress, which is expected since all internal forces must cancel. To this end, we observe that
\begin{equation}
\begin{split}
  & \nabla_{\alpha}\sum_{i=1}^N\sum_{j\neq i} f^{\text{int}}(\phi_i,\phi_j,\vec{\nabla}\phi_i,\vec{\nabla}\phi_j) \\
  =\;& 2\sum_{i=1}^N\sum_{j\neq i}\left(\frac{\pd f_{ij}^{\text{int}}}{\pd\phi_i}\nabla_{\alpha}\phi_i + \frac{\pd f_{ij}^{\text{int}}}{\pd\nabla_{\beta}\phi_i}\nabla_{\alpha}\nabla_{\beta}\phi_i\right) \;,
\end{split}
\end{equation}
where we have exploited the symmetry of $f_{ij}^{\text{int}}$. Hence, summing over all pairwise interactions leads to
\begin{equation}
\begin{split}
   &\sum_{i=1}^N\sum_{j\neq i}\left(\frac{\pd f_{ij}^{\text{int}}}{\pd\phi_i}-\nabla_{\beta}\frac{\pd f_{ij}^{\text{int}}}{\pd\nabla_{\beta}\phi_i}\right)\nabla_{\alpha}\phi_i \\
   =\;& \nabla_{\beta}\sum_{i=1}^N\sum_{j\neq i}\left(\delta_{\alpha\beta}\frac{f_{ij}^{\text{int}}}{2} - \frac{\pd f_{ij}^{\text{int}}}{\pd\nabla_{\beta}\phi_i}\nabla_{\alpha}\phi_i\right) \;,
\end{split}
\end{equation}
and so the total passive thermodynamic force density on the tissue can be written as
\begin{equation}
   \sum_{i=1}^N \rF_{i,\alpha}^{\text{pas}} \equiv \nabla_{\beta}\Sigma_{\alpha\beta}^{\text{pas}} \;,
\end{equation}
where 
\begin{equation}
\begin{split}
  \Sigma_{\alpha\beta}^{\text{pas}} =\;& \sum_{i=1}^N\Bigg[-\delta_{\alpha\beta}\Bigg(\Pi_i-\sum_{j\neq i}\frac{f_{ij}^{\text{int}}}{2}\Bigg) \\
  &\qquad\;\, -\Bigg(\frac{\pd f_{i}^{\text{CH}}}{\pd\nabla_{\beta}\phi_i}+\sum_{j\neq i}\frac{\pd f_{ij}^{\text{int}}}{\pd\nabla_{\beta}\phi_i}\Bigg)\nabla_{\alpha}\phi_i\Bigg] 
\end{split}  
\end{equation}
is the thermodynamic stress on the tissue. It should be noted that the divergence of the summand in this equation does not give the same result as $\rF_{i,\alpha}^{\text{pas}} = -\phi_i\nabla_{\alpha}\mu_i$ due to the interaction terms. Finally, substituting $f_{ij}^{\text{int}} = \epsilon\phi_i^2\phi_j^2$, we arrive at
\begin{equation}
\begin{split}
  \Sigma_{\alpha\beta}^{\text{pas}} =\;& \sum_{i=1}^N\Bigg[-\delta_{\alpha\beta}\Bigg(\Pi_i - \frac{\epsilon}{2}\sum_{j\neq i}\phi_i^2\phi_j^2\Bigg) \\
  &\qquad\;\, - 2\kappa\xi^2(\nabla_{\alpha}\phi_i)(\nabla_{\beta}\phi_i)\Bigg] \;,
\end{split}
\end{equation}
which is the result stated in \eqn\eqref{eqn:3D_passive_stress}.

\section{$z$-average of powers of the phase field}
\label{app:z_average_phi}

The goal of this section is to explicitly calculate the $z$-average of the assumed height profile of the phase field, which is given by
\begin{equation}
  \phi_{\ocmp}(z) = \frac{1}{2}\left[1+\tanh\left(\frac{h-z}{\xi}\right)\right]\,.
\end{equation}
Specifically, we show that in the limit where the interface thickness $\xi$ of the cell is much smaller than its height $h$ (i.e., $\vepsi \equiv \xi/h \ll 1$), one can write the average of this profile to any arbitrary power $n$ as
\begin{equation}
  \oavg{\phi_{\ocmp}^n} = 1 + k(n)\,\vepsi + \O\left(e^{-2/\vepsi}\right) \;, 
\end{equation}
where $k$ is a constant to be determined that solely depends on $n$. This result is used when evaluating $z$-averages of various quantities (e.g., the chemical potential $\mu$), both in the single-cell and multicellular cases. To verify this result, we first compute an exact expression for the $z$-average of $\phi_{\ocmp}^n$. This integral can be evaluated as
\begin{equation}
\begin{split}
  \oavg{\phi_{\ocmp}^n} &= \frac{1}{2^nh}\int_0^h\fd z\,\left[1+\tanh\left(\frac{h-z}{\xi}\right)\right]^n \\
  &= \frac{\vepsi}{2}\int_{\frac{1}{2}}^{g(\vepsi)}\fd\phi_{\ocmp}\,\frac{\phi_{\ocmp}^{n-1}}{1-\phi_{\ocmp}} \\
  &= \frac{\vepsi}{2}\left[B\left(g(\vepsi);n,0\right)-B\left(\frac{1}{2};n,0\right)\right] \;,
\label{eqn:avg_phi_z^n}
\end{split} 
\end{equation}
where $g(\vepsi) = \frac{1}{2}\left[1+\tanh\left(\frac{1}{\vepsi}\right)\right]$ and
\begin{equation}
  B(z;a,b) = \int_0^z\fd t\,t^{a-1}(1-t)^{b-1}
\end{equation}
is the incomplete Beta function. This function can be represented as a series:
\begin{equation}
  B(z;a,b) = z^a\sum_{m=0}^{\infty}\frac{(1-b)_m}{m!(a+m)}z^m \;,
\end{equation}
where $(\cdot)_m$ denotes the Pochhammer symbol [i.e., rising factorial with $(1)_m = m!$]. Next, to determine the behavior of $\oavg{\phi_{\ocmp}^n}$ when $\vepsi \ll 1$, we observe that
\begin{equation}
\begin{split}
    B(g(\vepsi);n,0) &= \sum_{m=0}^{\infty}\frac{g^{n+m}(\vepsi)}{n+m} = \sum_{m=n}^{\infty}\frac{g^m(\vepsi)}{m} \\
    &= -\log(1-g(\vepsi)) - \sum_{m=1}^{n-1}\frac{g^m(\vepsi)}{m} \;.
\end{split}
\end{equation}
By defining $\delta = 1-g(\vepsi)$ and using the binomial theorem, we have
\begin{align}
\begin{split}
  B(g(\vepsi);n,0) 
  =\;& -\log(\delta) - \sum_{m=1}^{n-1}\frac{1}{m} \\
  & - \sum_{m=1}^{n-1}\sum_{k=1}^m\frac{(-1)^k}{m}\binom{m}{k}\delta^k \;.
\end{split}
\end{align}
Noticing that $\sum_{m=1}^n\frac{1}{m} = H_n$ is the $n$th harmonic number and that the last sum on the right-hand side of this equation is simply a polynomial in $\delta$ of degree $n-1$, we can write
\begin{align}
  B(g(\vepsi);n,0) &= -\log(\delta) - H_{n-1} - \sum_{m=1}^{n-1} C_m\delta^m
\end{align}
for some finite coefficients $C_m$. Now when $\vepsi \ll 1$, we find
\begin{align}
  \delta = \frac{1}{2}\left[1-\tanh\left(\frac{1}{\vepsi}\right)\right] = \frac{1}{1+e^{2/\vepsi}} \approx e^{-2/\vepsi} \;,
\end{align}
and so
\begin{align}
  B(g(\vepsi);n,0) = \frac{2}{\vepsi} - H_{n-1} - \sum_{m=1}^{n-1}C_me^{-2m/\vepsi} \;.
\end{align}
Substituting this back to \eqn\eqref{eqn:avg_phi_z^n}, we arrive at the result stated at the beginning of this section:
\begin{align}
\begin{split}
  \oavg{\phi_{\ocmp}^n} &= \frac{\vepsi}{2}\left[\frac{2}{\vepsi} - H_{n-1} - B\left(\frac{1}{2};n,0\right) + \O\left(e^{-2/\vepsi}\right)\right] \\
    &= 1 + k(n)\,\vepsi + \O\left(e^{-2/\vepsi}\right) \;,
\end{split}
\end{align}
with
\begin{align}
  k(n) = -\frac{1}{2}\left[H_{n-1} + B\left(\frac{1}{2};n,0\right)\right] \;.
\end{align}

\section{Computing the $z$-average of\\ the passive stress for a single cell}
\label{app:single}

In the following, we compute explicitly the $z$-average of the in-plane components of the passive thermodynamic stress $\Sigma_{ab}^{\text{pas}}$ for a single cell, which is given by
\begin{equation}
\begin{split}
  \oavg{\Sigma}_{ab}^{\text{pas}} =\;& -\delta_{ab}\left(\oavg{\mu\phi}-\oavg{f^{\text{CH}}}+2\lambda\oavg{\delta V\phi^2}\right) \\
  &- 2\kappa\xi^2\oavg{(\nabla_{a}\phi)(\nabla_{b}\phi)} \;,
  \label{eqn:avg_passive_stress_single}
\end{split}
\end{equation}
where the chemical potential $\mu$ can be expressed as
\begin{equation}
  \mu = 2\kappa\left(2\phi^3-3\phi^2+\phi-\xi^2\nabla^2\phi\right) - 4\lambda\delta V\phi \;.
\end{equation}
We evaluate the averages in \eqn\eqref{eqn:avg_passive_stress_single} term by term. For the first term, we find 
\begin{equation}
\begin{split}
\oavg{\mu\phi} =\;& 2\kappa\phi_{\icmp}\big[2\phi_{\icmp}^3-3\phi_{\icmp}^2+\phi_{\icmp}-\xi^2\nabla_{\icmp}^2\phi_{\icmp} \\
& - \xi^2\oavg{\phi_{\ocmp}(\nabla_z^2\phi_{\ocmp})}\phi_{\icmp}\big] - 4\lambda\oavg{\delta V\phi_{\ocmp}^2}\phi_{\icmp}^2 + \O(\vepsi) \;,
\label{eq:mu2d}
\end{split}
\end{equation}
where we have used the result $\oavg{\phi_{\ocmp}^n}=1+\O(\vepsi)$ with $\vepsi = \xi/h$. To evaluate the remaining averages in this expression, we note that, for the given profile of $\phi_{\ocmp}$,
\begin{align}
  \xi\nabla_z\phi_{\ocmp} &= 2\phi_{\ocmp}(\phi_{\ocmp}-1) \;, \label{eq:dz_phiz} \\
  \xi^2\nabla_z^2\phi_{\ocmp} &=  4\phi_{\ocmp}(\phi_{\ocmp}-1)(2\phi_{\ocmp}-1) \;,
\end{align}
which are polynomials of $\phi_{\ocmp}$, and so
\begin{align}
  \xi^2\oavg{\phi_{\ocmp}\nabla_{z}^2\phi_{\ocmp}} = 4\left(2\oavg{\phi_{\ocmp}^4}-3\oavg{\phi_{\ocmp}^3}+\oavg{\phi_{\ocmp}^2}\right) = \O(\vepsi)\;.
\end{align}
Now since the cell volume can be expressed as
\begin{equation}
  V = \int_0^h\fd z\,\phi_{\ocmp}^2\int\fd\bx\,\phi_{\icmp}^2 = h\oavg{\phi_{\ocmp}^2}\,A \;,
\end{equation}
by defining $V_0 = A_0h$, we can write
\begin{equation}
  \delta V = 1 - \frac{V}{V_0} = 1 - \frac{A}{A_0}\oavg{\phi_{\ocmp}^2} = \delta A + \frac{A}{A_0}\left(1-\oavg{\phi_{\ocmp}^2}\right)\,,
\label{eq:3D_avg_vol}
\end{equation}
with $\delta A = 1-A/A_0$, and thus
\begin{equation}
  \oavg{\delta V \phi_{\ocmp}^2}  = \delta A + \O(\vepsi)\;.
\label{eq:avg_area_stress}
\end{equation}
Putting these results together, we obtain
\begin{equation}
  \oavg{\mu\phi} = \frac{\mu_{\icmp}\phi_{\icmp}}{h} + \O(\vepsi) \;,
  \label{eq:avg_phi_mu}
\end{equation}
with
\begin{equation}
\begin{split}
  \mu_{\icmp} = \;& 2\kappa_{\icmp}\left[\phi_{\icmp}(\phi_{\icmp}-1)(2\phi_{\icmp}-1)-\xi^2\nabla_{\icmp}^2\phi_{\icmp}\right] \\
  & -4\lambda_{\icmp}\delta A\phi_{\icmp}
\end{split}
\end{equation}
being the exchange chemical potential derived from a 2D free energy 
\begin{equation}
\begin{split}
  \F_{\icmp} =\;& \kappa_{\icmp}\int\fd\bx\,\left[f(\phi_{\icmp})+\xi^2(\vec{\nabla}_{\icmp}\phi_\icmp)^2\right] \\ 
  & + \lambda_{\icmp} A_0\,\delta A^2 \;,
\end{split}
\end{equation}
where $\kappa_{\icmp} = \kappa h$ and $\lambda_{\icmp} = \lambda h$.

The second term in the expression for $\oavg{\Sigma}_{ab}^{\text{pas}}$ can be computed as
\begin{align}
  \oavg{f^{\text{CH}}} &= \kappa\left[\oavg{f(\phi)} + \xi^2(\vec{\nabla}_{\icmp}\phi_{\icmp})^2\oavg{\phi_{\ocmp}^2} + \xi^2\phi_{\icmp}^2\oavg{(\nabla_z\phi_{\ocmp})^2}\right] \notag \\
  &= \frac{\kappa_{\icmp}}{h}\left[f(\phi_{\icmp}) + \xi^2(\vec{\nabla}_{\icmp}\phi_{\icmp})^2\right] + \O(\vepsi)\;,
  \label{eqn:avg_f_CH}
\end{align}
where we have used the fact that $\oavg{f(\phi)} = f(\phi_{\icmp}) + \O(\vepsi)$ and the result [see \eqn\eqref{eq:dz_phiz}]
\begin{equation}
  \xi^2\oavg{\left(\nabla_z\phi_{\ocmp}\right)^2} = 4\left(\oavg{\phi_{\ocmp}^4}-2\oavg{\phi_{\ocmp}^3}+\oavg{\phi_{\ocmp}^2}\right) = \O(\vepsi) \;,
\end{equation}

For the third and fourth averages, we have
\begin{equation}
  \oavg{\delta V\phi^2} = \oavg{\delta V\phi_{\ocmp}^2}\phi_{\icmp}^2 = \delta A \phi_{\icmp}^2 + \O(\vepsi)
  \label{eqn:avg_vol_phi2}
\end{equation}
and 
\begin{equation}
\begin{split}
  \oavg{\left(\nabla_a\phi\right)\left(\nabla_b\phi\right)} &= \oavg{\phi_{\ocmp}^2}\left(\nabla_a\phi_{\icmp}\right)\left(\nabla_b\phi_{\icmp}\right) \\
  &= \left(\nabla_a\phi_{\icmp}\right)\left(\nabla_b\phi_{\icmp}\right) + \O(\vepsi) \;.
  \label{eqn:avg_2d_grad_phi}
\end{split}
\end{equation}
Combining \eqns\eqref{eq:avg_phi_mu}, \eqref{eqn:avg_f_CH}, \eqref{eqn:avg_vol_phi2}, and \eqref{eqn:avg_2d_grad_phi}, we find
\begin{equation}
\begin{split}
  h\oavg{\Sigma}_{ab}^{\text{pas}} =\;& -\delta_{ab}\Pi_{\icmp} - 2\kappa_{\icmp}\xi^2\left(\nabla_a\phi_{\icmp}\right)\left(\nabla_b\phi_{\icmp}\right) \\
  &+ \O(\vepsi) \;,
  \label{eq:single_stress_avg_a}
\end{split}
\end{equation}
with $\Pi_{\icmp} = \mu_{\icmp}\phi_{\icmp}-f^{\text{CH}}(\phi_{\icmp}) + 2\lambda_{\icmp}\delta A\phi_{\icmp}^2$, as in \eqn\eqref{eqn:2D_passive_stressg_single} in the main text.

Finally, we show that this passive thermodynamic stress has a vanishing contribution to the shear stress acting on the cell along the interface with the substrate. Using the relation
\begin{equation}
\begin{split}
  \xi^2\nabla_z\phi_{\ocmp}|_{z=0} &= \frac{\xi}{2}\left[\tanh^2\left(\frac{1}{\vepsi}\right) - 1\right] \\
  &= \frac{\xi}{2}\left[\left(1-4e^{-2/\vepsi} + \dots\right) - 1\right] \\
  &= \O\left(e^{-2/\vepsi}\right) \;,
  \label{eq:xi2_dz_phiz_bc}
\end{split}
\end{equation}
we find
\begin{equation}
\begin{split}
  \Sigma_{az}^{\text{pas}}(\bx,z=0) &= - 2\kappa\xi^2\phi(\nabla_a\phi_{\icmp})(\nabla_z\phi_{\ocmp})|_{z=0} \\
  &= \O\left(e^{-2/\vepsi}\right) \;.
\end{split}
\end{equation}
As a result, the traction is solely determined by the stresses resulting from active processes that drive self-propulsion, i.e.,
\begin{equation}
  \rt_a(\bx) = \Sigma_{az}^{\text{pol}}(\bx,z=0) + \O\left(e^{-2/\vepsi}\right) \;.
\end{equation}

\section{Computing the $z$-averages of stresses for multiple cells in a cell monolayer}
\label{app:multi_average}

In this section, we compute the $z$-averages of those stresses that we include in the force balance equation for multiple cells. First, we consider the average of the in-plane passive thermodynamic stress, which can be expressed as (see \app\ref{app:passive_stress})
\begin{equation}
\begin{split}
  \oavg{\Sigma}_{ab}^{\text{pas}} = \sum_{i=1}^N\Bigg[& -\delta_{ab}\Bigg(\oavg{\Pi}_i-\frac{\epsilon}{2}\sum_{j\neq i}\oavg{\phi_i^2\phi_j^2}\Bigg) \\
  &- 2\kappa\xi^2\oavg{(\nabla_a\phi_i)(\nabla_b\phi_i)}\Bigg] \;,
\end{split}
\end{equation}
where $\oavg{\Pi_i} = \oavg{\mu_i\phi_i}-\oavg{f^{\text{CH}}(\phi_i)}+2\lambda\oavg{\delta V_i\phi_i^2}$. All $z$-averages within the summand can be computed using a similar procedure to that outlined in \app\ref{app:single} for the single-cell case [see \eqns\eqref{eq:avg_phi_mu}, \eqref{eqn:avg_f_CH}, \eqref{eqn:avg_vol_phi2}, and \eqref{eqn:avg_2d_grad_phi}]. The only exception is the average of the interaction term, which can be evaluated as 
\begin{equation}
  \oavg{\phi_i^2\phi_j^2} = \oavg{\phi_{\ocmp}^4}\phi_{i\icmp}^2\phi_{j\icmp}^2 = \phi_{i\icmp}^2\phi_{j\icmp}^2 + \O(\vepsi) \;,
\end{equation}
where we have used the result $\oavg{\phi_{\ocmp}^n} = 1 + \O(\vepsi)$. Hence,
\begin{equation}
\begin{split}
  h\oavg{\Sigma}_{ab}^{\text{pas}} = \sum_{i=1}^N &\Bigg[ -\delta_{ab}\Bigg(\Pi_{i\icmp} - \frac{\epsilon_{\icmp}}{2}\sum_{j\neq i}\phi_{i\icmp}^2\phi_{j\icmp}^2 \Bigg) \\
  &- 2\kappa_{\icmp}\xi^2(\nabla_a\phi_{i\icmp})(\nabla_b\phi_{i\icmp})\Bigg] + \O(\vepsi) \;,
\end{split}
\end{equation}
with $\Pi_{i\icmp} = \mu_{i\icmp}\phi_{i\icmp}-f^{\text{CH}}(\phi_{i\icmp})+2\lambda_{\icmp}\delta A_i\phi_{i\icmp}^2$. Here, $\mu_{i\icmp}$ is a 2D chemical potential for cell $i$ given by
\begin{equation}
\begin{split}
  \mu_{i\icmp} =\;& 2\kappa_{\icmp}\left[\phi_{i\icmp}(\phi_{i\icmp}-1)(2\phi_{i\icmp}-1)-\xi^2\nabla_{\icmp}^2\phi_{i\icmp}\right] \\
  & -4\lambda_{\icmp}\delta A_i\phi_{i\icmp} + 2\epsilon_{\icmp}\phi_{i\icmp}\sum_{j\neq i}\phi_{j\icmp}^2 \;,
\end{split}
\end{equation}
where $\epsilon_{\icmp} = \epsilon h$. Since this 2D stress has the same form as the 3D one, we can therefore take advantage of the theory developed in \app\ref{app:passive_stress} and write
\begin{equation}
  h\nabla_b\oavg{\Sigma}_{ab}^{\text{pas}} = -\sum_{i=1}^N \phi_{i\icmp}\nabla_a\mu_{i\icmp} \;,
\end{equation}
which is \eqn\eqref{eqn:2D_passive_stress} in the main text. 

We also claim that $\Sigma_{az}^{\text{pas}}(\bx,z=0) \approx 0$. Using the result $\xi^2\nabla_z\phi_{\ocmp}|_{z=0} = \O\left(e^{-2/\vepsi}\right)$ from \eqn\eqref{eq:xi2_dz_phiz_bc}, we obtain
\begin{align}
  \Sigma_{az}^{\text{pas}}(\bx,z=0) &= -2\kappa\xi^2\sum_{i=1}^N\phi_i(\nabla_a\phi_{i\icmp})(\nabla_z\phi_{\ocmp})|_{z=0} \notag \\
  &= \O\left(e^{-2/\vepsi}\right) \;,
\end{align}
which is consistent with our claim when $\vepsi$ is small.

To average the in-plane viscous stress $\Sigma_{ab}^{\text{vis}}$, we observe that
\begin{equation}
  \oavg{\rV}_a = \rV_{\icmp,a}\oavg{\mathcal{V}(z)} = \rV_{\icmp,a}\oavg{\frac{z}{h^2}(2h-z)} = \frac{2}{3}\rV_{\icmp,a} \;,
\end{equation}
and thus
\begin{equation}
\begin{split}
  \oavg{\Sigma}_{ab}^{\text{vis}} =\;& \eta\left(\nabla_a\oavg{\rV}_b+\nabla_b\oavg{\rV}_a\right)+\left(\zeta-\frac{2}{3}\eta\right)\delta_{ab}\oavg{\nabla_{\gamma}\rV_{\gamma}} \\
  =\;& \frac{1}{h}\Bigg[\eta_{\icmp} \left(\nabla_a\rV_{\icmp,b}+\nabla_b\rV_{\icmp,a}\right) \\
  & \quad +\left(\zeta_{\icmp}-\frac{2}{3}\eta_{\icmp}\right)\delta_{ab}\nabla_c\rV_{\icmp,c}\Bigg] \;,
\end{split}
\end{equation}
where we have used the fact that $\rV_{\ocmp} = 0$ and defined $\eta_{\icmp} = 2\eta h/3$ and $\zeta_{\icmp} = 2\zeta h/3$. Taking the divergence, we then find
\begin{equation}
  h\nabla_{b}\oavg{\Sigma}_{ab}^{\text{vis}} = \eta_{\icmp}\nabla_b\nabla_b\rV_{\icmp,a} + \nu_{\icmp}\nabla_a\nabla_b\rV_{\icmp,b} \;,
\end{equation}
with $\nu_{\icmp} = \zeta_{\icmp}+\eta_{\icmp}/3$, thereby recovering \eqn\eqref{eqn:2D_viscous_stress} in the main text. 

Finally, we note that this viscous stress contributes to the traction upon averaging over $z$. This is because
\begin{equation}
  \Sigma_{az}^{\text{vis}}(\bx,z=0) = \eta\nabla_z\rV_a|_{z=0} = \frac{3\eta_{\icmp}}{h^2}\rV_{\icmp,a} \;,
\end{equation}
which is the result stated in \eqn\eqref{eq:viscous_traction}.

\section{Projection of the averaged\\ in-plane viscous stress}
\label{app:project}

This appendix explains how the viscous stress $\oavg{\Sigma}_{ab}^{\text{vis}}$ at the tissue level can give rise to pairwise friction at the cellular level. First, recall that we define the viscous body force acting on a cell (say cell $i$) by projecting the viscous force density at the tissue level $\rF_a^{\text{vis}} = h\nabla_b\oavg{\Sigma}_{ab}^{\text{vis}}$ onto $\phi_{i\icmp}$ as follows:
\begin{widetext}  
\begin{equation}
  \rf_{i,a}^{\text{vis}} = \int\fd\bx\,\phi_{i\icmp}\rF_a^{\text{vis}} = \int\fd\bx\,\phi_{i\icmp}\left(\eta_{\icmp}\nabla_b\nabla_b\rV_{\icmp,a}+\nu_{\icmp}\nabla_a\nabla_b\rV_{\icmp,b}\right) \;.
\end{equation}
We evaluate this integral term by term. Integrating by parts and observing that
\begin{equation}
  \nabla_b\rV_{\icmp,a} = 
  \frac{1}{\Phi_{\icmp}}\sum_{j=1}^N(\nabla_b\phi_{j\icmp})\rv_{j,a}^{\text{c}}
  - \frac{1}{\Phi_{\icmp}^2}\sum_{j=1}^N\phi_{j\icmp}(\nabla_b\Phi_{\icmp})\rv_{j,a}^{\text{c}} 
  = \frac{1}{\Phi_{\icmp}}\sum_{j=1}^N(\nabla_b\phi_{j\icmp})\left(\rv_{j,a}^{\text{c}}-\rV_{\icmp,a}\right) \;,
\end{equation}
we find
\begin{equation}
  \int\fd\bx\,\phi_{i\icmp}\nabla_b\nabla_b\rV_{\icmp,a}
  = \sum_{j=1}^N\int\fd\bx\,\frac{1}{\Phi_{\icmp}}D_{ij,bb}\left(\rV_{\icmp,a}-\rv_{j,a}^{\text{c}}\right) \;,
\end{equation}
and
\begin{equation}
  \int\fd\bx\,\phi_{i\icmp}\nabla_a\nabla_b\rV_{\icmp,b}
  = \sum_{j=1}^N\int\fd\bx\,\frac{1}{\Phi_{\icmp}}D_{ij,ab}\left(\rV_{\icmp,b}-\rv_{j,b}^{\text{c}}\right)
\end{equation}
where $D_{ij,ab} = (\nabla_a\phi_{i\icmp})(\nabla_b\phi_{j\icmp})$. Combining the results from these two integrals gives
\begin{equation}
  \rf_{i,a}^{\text{vis}} = \sum_{j=1}^N\int\fd\bx\,M_{ij,ab}\left(\rV_{\icmp,b}-\rv_{j,b}^{\text{c}}\right) \;,
  \label{eq:viscous_force_projection_app}
\end{equation}
with
\begin{equation}
  M_{ij,ab} = \frac{1}{\Phi_{\icmp}}\left(\eta_{\icmp}D_{ij,cc}\delta_{ab} +\nu_{\icmp}D_{ij,ab} \right) \;,
  \label{eq:viscous_interface_matrix_app}
\end{equation}
which correspond to \eqns\eqref{eq:viscous_force_projection} and \eqref{eq:viscous_interface_matrix} in the main text (in component form). We now consider the limit when three-body overlaps are rare and neglect terms in \eqn\eqref{eq:viscous_force_projection_app} that are products of three different phase fields and/or their gradients. In that equation, we confront with terms of the form
\begin{equation}
  \sum_{j=1}^N D_{ij,ab}\left(\rV_{\icmp,b}-\rv_{j,b}^{\text{c}}\right) = \frac{1}{\Phi_{\icmp}}\sum_{j=1}^N\sum_{k=1}^N D_{ij,ab}\,\phi_{k\icmp}\left(\rv_{k,b}^{\text{c}}-\rv_{j,b}^{\text{c}}\right) \;.
\end{equation}
Keeping only terms that involve less than three different phase fields, we find
\begin{equation}
  \sum_{j=1}^N\sum_{k=1}^N D_{ij,ab}\,\phi_{k\icmp}\rv_{k,b}^{\text{c}}
  \approx \sum_{j=1}^N D_{ij,ab}\,\phi_{j\icmp}\rv_{j,b}^{\text{c}} +
  \sum_{j\neq i} D_{ii,ab}\,\phi_{j\icmp}\rv_{j,b}^{\text{c}} +
  \sum_{j\neq i} D_{ij,ab}\,\phi_{i\icmp}\rv_{i,b}^{\text{c}} \;\;
  \label{eq:viscous_interface_vk}
\end{equation}
and 
\begin{equation}
  \sum_{j=1}^N\sum_{k=1}^N D_{ij,ab}\,\phi_{k\icmp}\rv_{j,b}^{\text{c}}
  \approx \sum_{j=1}^N D_{ij,ab}\,\phi_{j\icmp}\rv_{j,b}^{\text{c}} +
  \sum_{j\neq i} D_{ii,ab}\,\phi_{j\icmp}\rv_{i,b}^{\text{c}} +
  \sum_{j\neq i} D_{ij,ab}\,\phi_{i\icmp}\rv_{j,b}^{\text{c}} \;.
  \label{eq:viscous_interface_vj}
\end{equation}
Subtracting \eqn\eqref{eq:viscous_interface_vk} by \eqn\eqref{eq:viscous_interface_vj}, we can then write
\begin{equation}
  \sum_{j=1}^N\sum_{k=1}^N D_{ij,ab}\,\phi_{k\icmp}\left(\rv_{k,b}^{\text{c}}-\rv_{j,b}^{\text{c}}\right) \approx \sum_{j\neq i} \mathcal{D}_{ij,ab}\left(\rv_{i,b}^{\text{c}}-\rv_{j,b}^{\text{c}}\right) \;,
\end{equation}
where we defined
\begin{equation}
  \mathcal{D}_{ij,ab} = D_{ij,ab}\,\phi_{i\icmp}-D_{ii,ab}\,\phi_{j\icmp} \;.
\end{equation}
Applying this relation to the terms in \eqn\eqref{eq:viscous_force_projection_app}, we arrive at
\begin{equation}
  \rf_{i,a}^{\text{vis}} \approx \sum_{j\neq i}\int\fd\bx\,\widetilde{M}_{ij,ab}\left(\rv_{i,b}^{\text{c}}-\rv_{j,b}^{\text{c}}\right) \;,
\end{equation}
which is the result stated in the main text, with
\begin{equation}
    \widetilde{M}_{ij,ab} = \frac{1}{\Phi_{\icmp}^2}\left(\eta_{\icmp}\mathcal{D}_{ij,cc}\delta_{ab} + \nu_{\icmp}\mathcal{D}_{ij,ab}\right) \;.
\end{equation}
\end{widetext}

\begin{table*}[!th]
\begin{ruledtabular}
\begin{tabular}{cccc}
\textrm{Parameter}&
\textrm{Interpretation}&
\textrm{Dimensions}&
\textrm{Value(s)}\\
\colrule
  $d$ & Deformability & - & $0.25$--$4.16$ \\
  $R$ & Ideal cell radius & [L] & $1$ \\
  $\xi$ & Cell interface thickness & [L] & $1/8$ \\
  $\epsilon_{\icmp}$ & Cell-cell repulsion & [E][L]$^{-2}$ & $1$ \\
  $\omega_{\icmp}$ & Cell-cell adhesion & [E][L]$^{-2}$ & $0$ or $0.065$ \\
  $\lambda_{\icmp}$ & Cell area constraint & [E][L]$^{-2}$ & $1$ \\
  $\eta_{\icmp}$ & Cell-cell friction & [E][T][L]$^{-2}$ & 0, $0.0016$, or $0.0064$ \\
  $\Gamma^{\phi}$ & Inverse mobility & [E][T][L]$^{-2}$ & $0.01$ \\
  $\Gamma_{\icmp}$ & Cell-substrate friction & [E][T][L]$^{-4}$ & $0.01$ \\
  $v_0$ & Cell motility & [L][T]$^{-1}$ & $0.25$--$2.0$ \\
  $D_r$ & Rotational diffusion rate & [T]$^{-1}$ & $1$ \\
  $\delta t$ & Timestep & [T] & $10^{-3}$ \\
  $\delta x$ & Lattice size & [L] & $1/8$ \\
  $L_{\text{sub}}$ & Cell subdomain size & [L] & $4.75$ \\
  $L$ & Simulation box size & [L] & $18.125$ \\
  $\varphi$ & Packing fraction & - & $0.96$ \\
\end{tabular}
\end{ruledtabular}
\caption{\label{tab:sim_params}
Parameter value(s) used in the simulations to determine the phase diagrams (\fig\ref{fig:pd}) and to compare the effect on velocity correlations of the cells between cell-cell adhesion and intercellular friction (\fig\ref{fig:vel_cor}).}
\end{table*}

\section{Simulation details}
\label{app:sims}

We report the simulation parameter values in Table~\ref{tab:sim_params}, with the preferred cell radius $R$ as the unit of length and the inverse of the rotational diffusion coefficient $D_r^{-1}$ as the unit of time. We also use the strength of steric repulsion $\epsilon_{\icmp}$ to characterize the energy scales. We employ finite differences to evolve the field equations, where we use fourth-order central finite differences for the gradient terms and a nine-point stencil for the Laplacian to ensure stability. We also use a third-order upwind scheme for stabilizing the advection velocities, which, for the values of $\delta t$ and $v_0$ we have considered,  remain small enough to avoid any instabilities. When including intercellular friction, we use \texttt{LAPACK} for solving the matrix equation [\eqn\eqref{eqn:cell_advection_matrix}] and, for simplicity, we set $\eta_{\icmp} = \nu_{\icmp}$. As in previous work, we parallelize the model by using an auxiliary field so that we can solve the individual phase fields on subdomains of dimensions $L_{\text{sub}} \times L_{\text{sub}}$, with fixed boundary conditions $\phi_i=0$ on the edge of the subdomain. We shift the position of the cell within its subdomain when it moves more than two lattice points in any direction (along with the position of the subdomain relative to the full simulation lattice), which allows the cell to remain localized to the center of its subdomain. We initialize the cells as circles, randomly positioned but at least $R/2$ apart in the full domain, with $\phi_i=1$ inside the cell and $\phi_i=0$ outside. We perform a passive run of $10^5$ timesteps to allow the system to equilibrate and reach confluence. We then turn on motility ($v_0 \neq 0$) for $10^6$ timesteps before taking data for an additional $10^7$ timesteps (for the data in \fig\ref{fig:vel_cor}, we only require an additional $10^6$ timesteps to determine the steady-state velocity correlation functions). We parallelize each simulation on 12 processors via \texttt{OpenMP}.

\end{document}